\documentclass[runningheads]{llncs}
\usepackage{graphicx}
\usepackage{subcaption}
\usepackage{xcolor}
\usepackage[page]{appendix}
\usepackage{array,multirow}
\usepackage{wrapfig}
\begin{document}

\title{Scouting the Path to a Million-Client Server\thanks{A version of this paper was published in the following conference paper: Zhao Y., Saeed A., Ammar M., Zegura E. (2021) Scouting the Path to a Million-Client Server. In: Hohlfeld O., Lutu A., Levin D. (eds) Passive and Active Measurement. PAM 2021. Lecture Notes in Computer Science, vol 12671. Springer, Cham. \url{https://doi.org/10.1007/978-3-030-72582-2_20}}}

\author{Yimeng Zhao \inst{1} \and Ahmed Saeed\inst{2},
Mostafa Ammar \inst{1} \and Ellen Zegura \inst{1}}

\institute{Georgia Institute of Technology
\and
Massachusetts Institute of Technology
}

\vspace{-1in}

\maketitle

\begin{abstract}
To keep up with demand, servers will scale up to handle hundreds of thousands of clients simultaneously. Much of the focus of the community has been on scaling servers in terms of aggregate traffic intensity (packets transmitted per second). However, bottlenecks caused by the increasing number of concurrent clients, resulting in a large number of concurrent flows, have received little attention. In this work, we focus on identifying such bottlenecks. In particular, we define two broad categories of problems;
namely, admitting more packets into the network stack than can be handled efficiently, and increasing per-packet overhead within the stack. We show that these problems contribute to high CPU usage and network performance degradation in terms of aggregate throughput and RTT. Our measurement and analysis are performed in the context of the Linux networking stack, the the most widely used publicly available networking stack. Further, we discuss the relevance of our findings to other network stacks. The goal of our work is to highlight considerations required in the design of future networking stacks to enable efficient handling of large numbers of clients and flows. 

\end{abstract}

\section{Introduction}

Modern servers at large scale operators handle tens of thousands of clients simultaneously \cite{acceltcp,zhang2017tuning,carousel}. This scale will only grow as NIC speeds increase \cite{100gbps_nic,8207825,ethernet_roadmap} and servers get more CPU cores \cite{geer2005chip,core_trend}. For example, a server with a 400 Gbps NIC \cite{8207825} can serve around 80k HD video clients and 133k SD video clients. \footnote{HD and SD videos consume up to 5 Mbps and 3 Mbps, respectively \cite{netflix_bitrate}.} 
This scale is critical not only for video on demand but also for teleconferencing and AR/VR applications. The focus of the community has been on scaling servers in terms of packets transmitted per second \cite{netmap,kalia2019datacenter,flexnic,hock2019tcp,shenango,ix}, with little attention paid to developing complete stacks that can handle large numbers of flows well \cite{mtcp,tas}. 

We envisage servers delivering large volumes of data to millions of clients simultaneously. Our goal is to identify bottlenecks that arise when servers reach that scale. In particular, we take a close look at network stack components that become the bottleneck as the number of flows increases. We find that competition between flows can lead to overall performance degradation, requiring fine-grain scheduling. Further, the increase in flow numbers leads to higher overhead of per-flow bookkeeping and flow coordination. Thus, we categorize problems that arise due to an increase in the number of concurrent flows into two categories: 

\noindent \textbf{1) Admission Control to the Stack:} The admission policy determines the frequency at which a flow can access the stack and how many packets it can send per access. The frequency of a flow accessing network resources and the duration of each access determine the throughput it can achieve. As the number of flows increases, admission control becomes critical for the efficiency of the stack. For example, admitting and alternating between flows at a high frequency can reduce Head-of-Line (HoL) blocking and improve fairness but at the expense of CPU overhead, which can become a bottleneck, leading to throughput loss. We consider backpressure mechanism as a critical part of the admission control as it determines how a flow is paused (e.g., denied admission) and resumed (i.e., granted admission).


\noindent \textbf{2) Per-packet Overhead within the Stack:} The overhead of most per-packet operations is almost constant or a function of packet size (e.g., checksum, routing, and copying). However, the overhead of some operations depends entirely on the number of flows serviced by the system. For example, the overhead of matching an incoming packet to its flow (i.e., demultiplexing), and the overhead of scheduling, for some scheduling policies (e.g., fair queueing), are tied to the number of flows in the system. 

\begin{table}[!t]
\begin{center}
\scriptsize
\begin{tabular}{|c| c | c | c |}
\hline
Category& Identified Issue & Impact & Existing systems mitigating it \\ 
\hline
\hline
\multirow{5}{*}{\rotatebox{90}{Admission} \rotatebox{90}{\hspace{0.05in}Control}} &  Overpacing & 5\% increase in CPU utilization & -\\
\cline{2-4}
&Inefficient  & Throughput unfairness and hundreds  & \multirow{2}{*}{Per-flow scheduling \cite{zd,snap}} \\
&backpressure & of milliseconds in latency & \\
\cline{2-4}
& Oblivious hardware  & \multirow{2}{*}{2$\times$ increase in interrupts} & \multirow{2}{*}{-}\\
& offloads & & \\
\hline
\multirow{5}{*}{\rotatebox{90}{Per-packet}\rotatebox{90}{\hspace{0.02in}Overhead}} &Data structure & 2$\times$ increase in CPU utilization  & Low-overhead data\\
& inefficiency & and  2$\times$ increase in latency & structures \cite{carousel,eiffel} \\
\cline{2-4}
&\multirow{2}{*}{Lock contention} & \multirow{2}{*}{2$\times$ increase in latency} & Distributed scheduling \\
& & &  \cite{carousel,stephens2017titan,multi_queue} \\
\cline{2-4}
&Cache pressure & 1.8$\times$ increase in latency & -\\
\hline
\end{tabular}
\end{center}
\vspace{-0.1in}
\caption{Summary of findings with results reported at 100k flows compared to more efficient baselines for admission control or performance with lower number of flows for for per-packet overhead.}
\label{tab:1}
\vspace{-0.4in}
\end{table}



We focus our attention on Linux servers. Despite its well documented inefficiencies (e.g., the overhead of system calls, interrupts, and per-packet memory allocation \cite{mtcp,brouer2015network}), the Linux networking stack remains the most widely used publicly available networking stack. Further, even when new userspace stacks are deployed, they still rely, at least partially, on the Linux stack to make use of its comprehensive Linux functionality and wide use \cite{snap}. Hence, our focus on Linux is critical for two reasons: 1) our results are immediately useful to a wide range of server operators, and 2) we are able to identify all possible bottlenecks that might not appear in other stacks because they lack the functionality. 

We focus on the overhead of long-lived flows. Long-lived flows help expose problems related to scaling a stack in terms of the number of flows. Scheduling long-lived flows requires the scheduler to keep track of all active flows, exposing inefficient data structures whose overhead increases with the number of tracked flows and highlighting issues that arise because of the interactions between the transport layer and the scheduler. It also exposes cache inefficiencies as information about a flow has to be retained and edited over a long period of time. Applications with long-lived
flows include video on demand and remote storage. The inefficiency of short-lived flows is rooted in creation and destruction of states, and has been studied in earlier work \cite{acceltcp}



{\em The contribution of this work is in evaluating the scalability of the network stack as a whole, at hundreds of thousands of clients, leading to the definition of broader categories of scalability concerns.} Table~\ref{tab:1} summarizes our findings and existing systems that mitigating the problems. It should be noted that inefficient backpressure and data structure problems are only partially addressed by the existing solutions and we'll discuss the remaining challenges in section \ref{sec:admission} and \ref{sec:packet-overhead}. In earlier work there have been several proposals to improve the scalability of different components of the network stack (e.g., transport layer \cite{mtcp,tas,acceltcp} and scheduling \cite{qfq,carousel,eiffel}). These proposals consider specific issues with little attempt to generalize or categorize such scalability concerns. Further, the notion of scalability considered in earlier work is still limited to tens of thousands of flows, with a general focus on short flows.






\section{Measurement Setup}

\textbf{Testbed:} We conduct experiments on two dual-socket servers. Each server is equipped with two Intel E5-2680 v4 @ 2.40GHz processors. Each server has an Intel XL710 Dual Port 40G NIC Card with multi-queue enabled. The machines belong to the same rack. Both machines use Ubuntu Server 18.04 with Linux kernel 5.3.0. 

\textbf{Testbed Tuning:} 
The affinity of the interrupts and application to CPU cores significantly affects the network performance on a multi-core and multi-socket machine. 
To reduce cache synchronization between different cores and improve interrupt affinity, we pin
each transmit/receive queue pair to the same core. We enable Receiver Packet Steering (RPS), which sends the packet to a CPU core based on the hash of source and destination IPs and ports. 
We limit all network processing to exclusively use the local socket because we observe that the interconnection between different sockets leads to performance degradation at 200k or more flows. 
We enabled different hardware offload functions including GSO, GRO, and LRO to lower CPU utilization. We also enabled interrupt moderation to generate interrupts per batch, rather than per packet. We use TCP CUBIC as the default transport protocol, providing it with maximum buffer size, to avoid memory bottlenecks. The entire set of parameters is shown in Appendix \ref{apx:config}.


\textbf{Traffic Generation:} We generate up to 300k concurrent flows with \verb|neper|~\cite{neper}. We bind multiple IP addresses to each server so the number of flows that can be generated is not limited by the number of ports available for a single IP address. With 40 Gbps aggregate throughput, the per-flow rate can range from 133 Kbps, which is a typical flow rate for web service \cite{cavalcanti2009optimizing}, to 400 Mbps, which might be large data transfer \cite{chen2011peer}.
We ran experiments with different numbers of threads ranging from 200 to 2000. In particular, we spawn N threads, create M flows that last for 100s, and multiplex the M flows evenly over the N threads. We observed that using more threads causes higher overhead in book-keeping and context switch, leading to degraded throughput when the server needs to support hundreds of thousands of flows. The results shown in this paper are with 200 threads if not specified otherwise. We use long-lived flows for experiments because our focus is on the scaling problem in terms of the number of concurrent flows. The scaling problem of short-lived flows is more related to the number of connecting requests per second rather than the number of concurrent flows. With fixed number of flows, the short-lived flows should not have higher overhead than long-lived flows. For the rest of the paper, we use flows and clients interchangeably.

\begin{figure}[!t]
\vspace{-0.1in}
\centering
\includegraphics[width=0.75\linewidth]{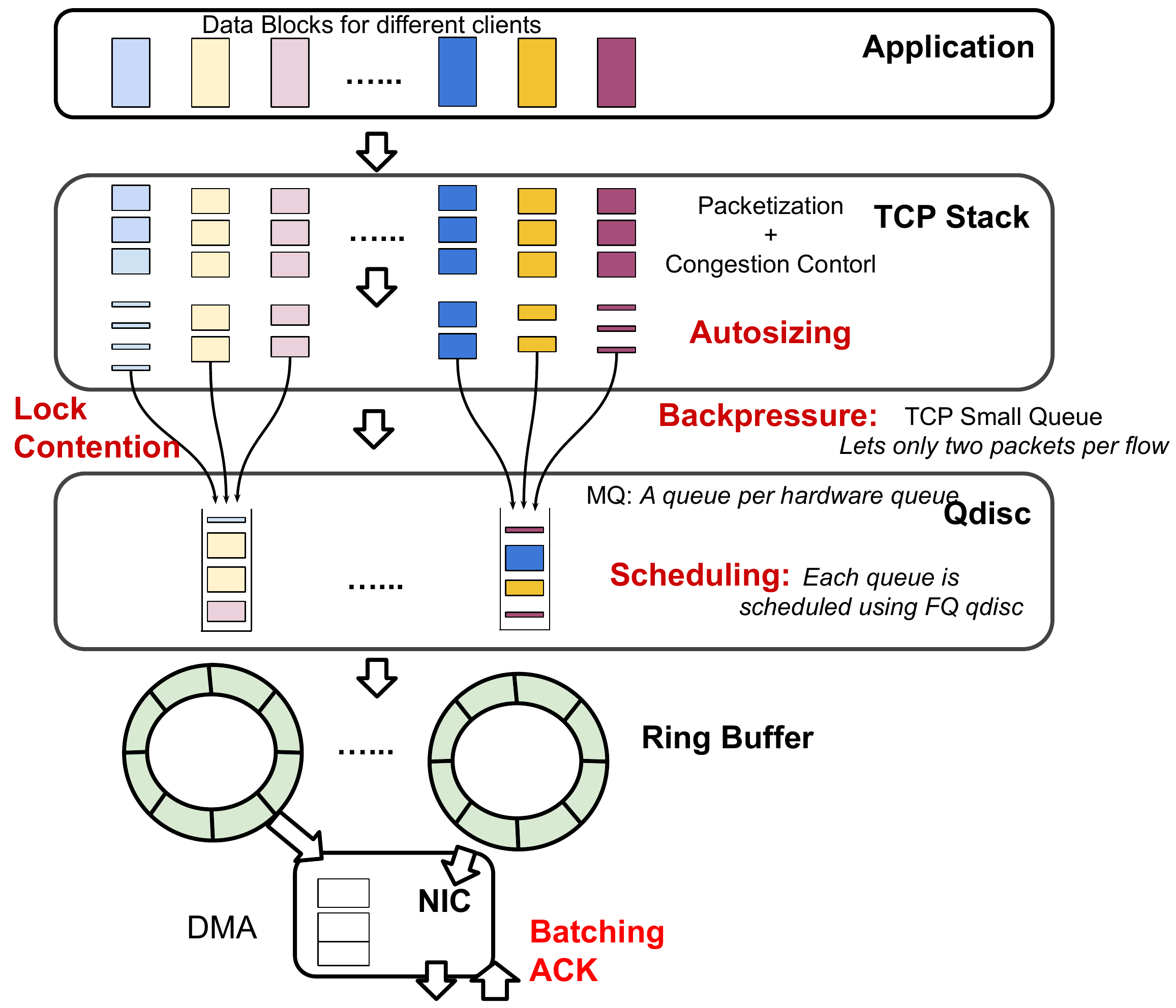}
\vspace{-0.1in}
\caption{Schematic of the packet transmission path with identified pain points marked in red.}
\label{fig:pkt-xmit-stacks}
\vspace{-0.24in}
\end{figure}

Figure \ref{fig:pkt-xmit-stacks} visualizes our assumed stack architecture. Our focus is on the overhead of the transport and scheduling components of the stack. We experiment with different scheduling algorithms by installing different Queuing Disciplines (qdiscs). We use multiqueue qdisc (\verb|mq|) to avoid having a single lock for all hardware queues. All scheduling algorithms are implemented by per-queue within \verb|mq|. 
By default, \verb|mq| handles packets FIFO in its queues. However, we use Fair Queue (\verb|fq|) \cite{fq} as the default qdisc combined with \verb|mq|. Compared to \verb|pfifo_fast|, \verb|fq| achieves better performance in terms of latency and CPU usage when handling a large number of flows \cite{zd}. In some experiments, we limit the total flow rate to 90\% of the link speed to avoid queueing in Qdiscs and show that the performance degradation cannot be avoided by simply lowering the total rate. We also use \verb|fq_codel| \cite{fq_codel} to reduce latency within the qdisc in some cases.






\textbf{Measurement Collection:} In all experiments, machines are running only the applications mentioned here making any CPU performance measurements correspond with packet processing. We track overall CPU utilization using \verb|dstat|~\cite{dstat-cmd} and track average flow RTT using \verb|ss|~\cite{ss-cmd}. We track the TCP statistics using \verb|netstat|~\cite{netstat-cmd}. Performance statistics of specific functions in the kernel is obtained using \verb|perf|~\cite{perf-cmd}.

\vspace{-0.1in}
\section{Overall Stack Performance}

\begin{wrapfigure}{r}{0.62\textwidth}
\vspace{-0.3in}
\centering
  \begin{subfigure}{0.31\textwidth}
    \centering
    \includegraphics[width=\linewidth]{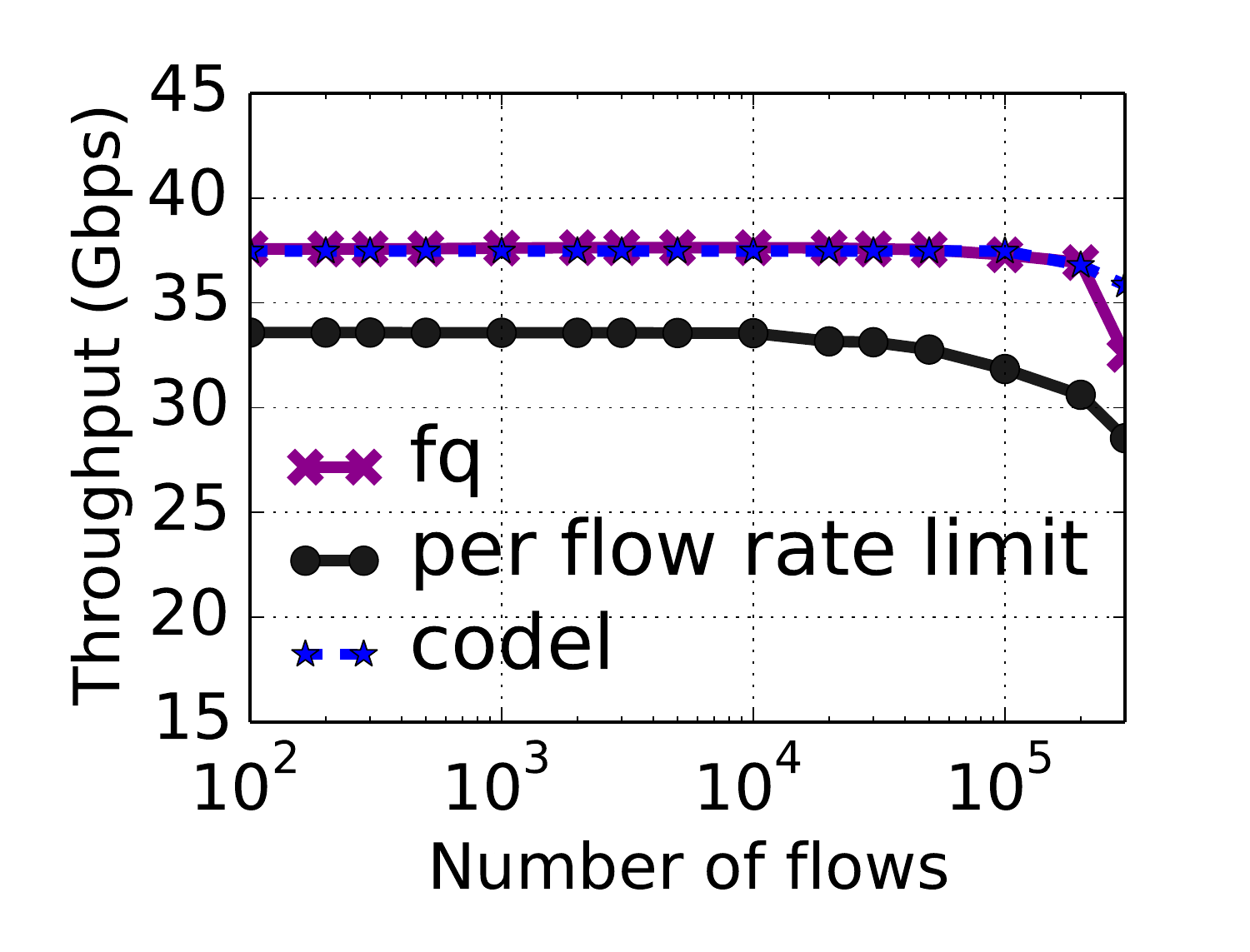}
    \vspace{-0.25in}
    \caption{\small Aggregate Throughput}
    \label{fig:summary-throughput}
    \end{subfigure}
    \begin{subfigure}{0.3\textwidth}
    \centering
    \includegraphics[width=\linewidth]{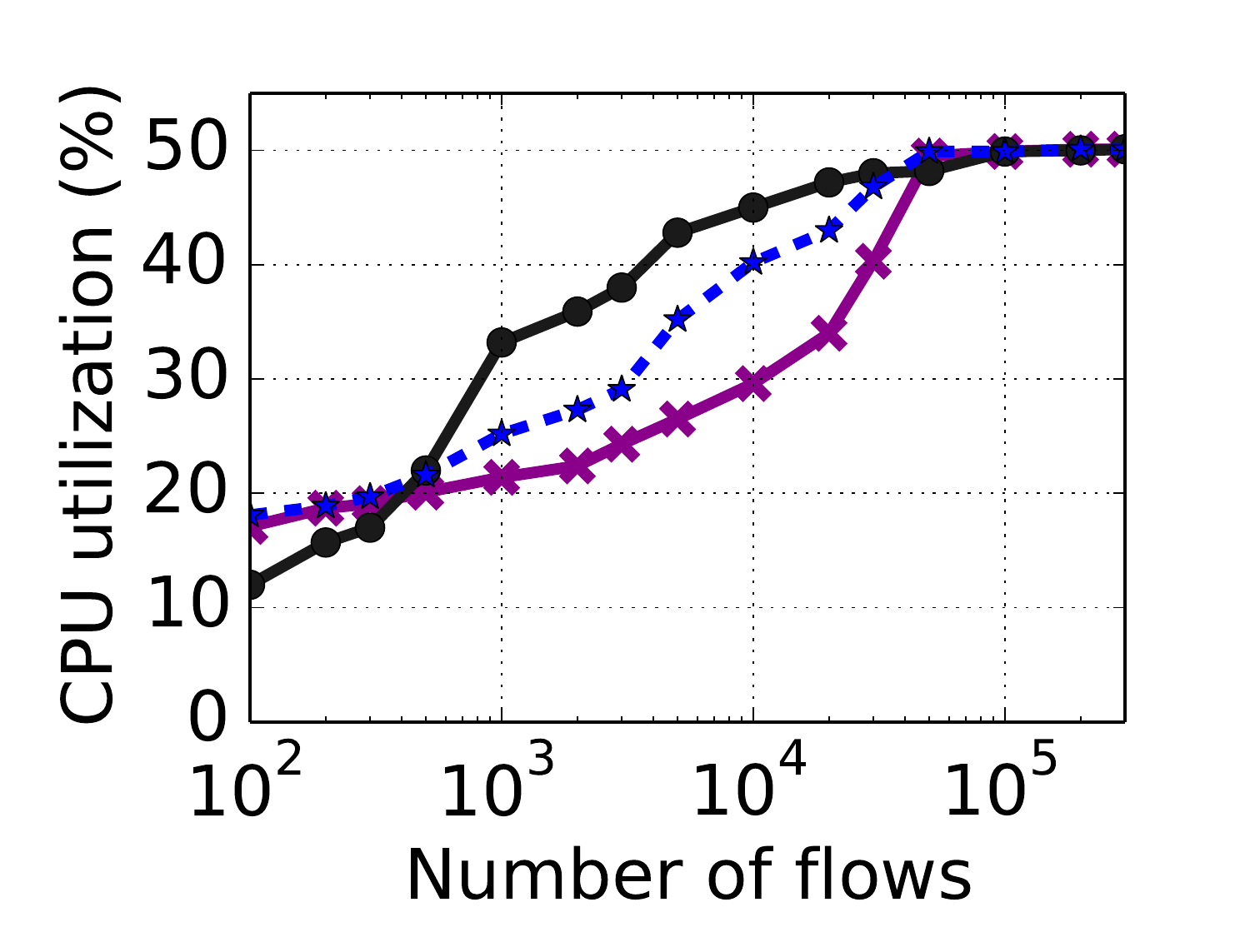}
    \vspace{-0.25in}
    \caption{\small CPU Usage}
    \label{fig:summary-cpu}
    \end{subfigure}
    \begin{subfigure}{0.31\textwidth}
    \centering
    \includegraphics[width=\linewidth]{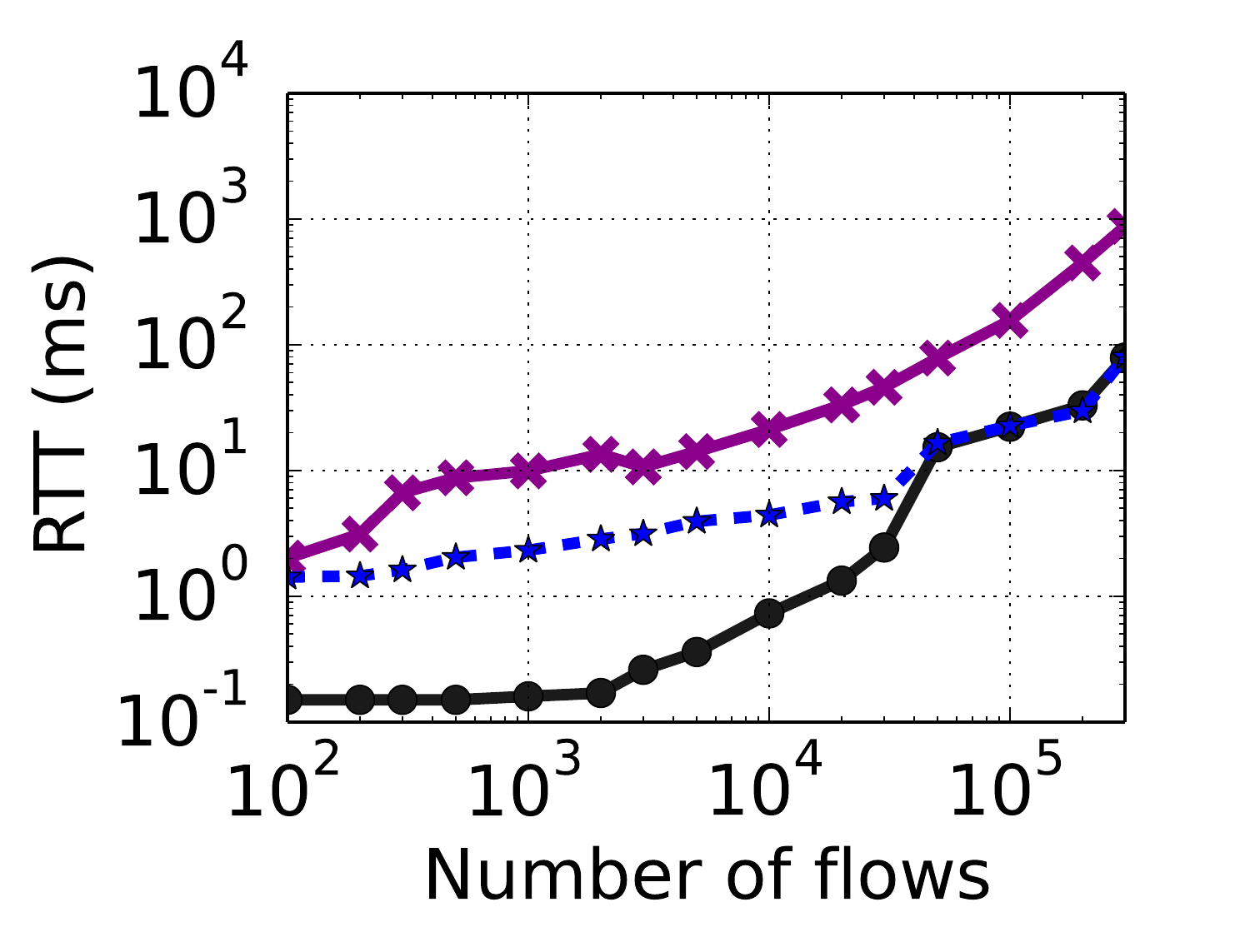}
    \vspace{-0.25in}
    \caption{\small RTT}
    \label{fig:summary-rtt}
    \end{subfigure}
    \begin{subfigure}{0.3\textwidth}
    \centering
    \includegraphics[width=\linewidth]{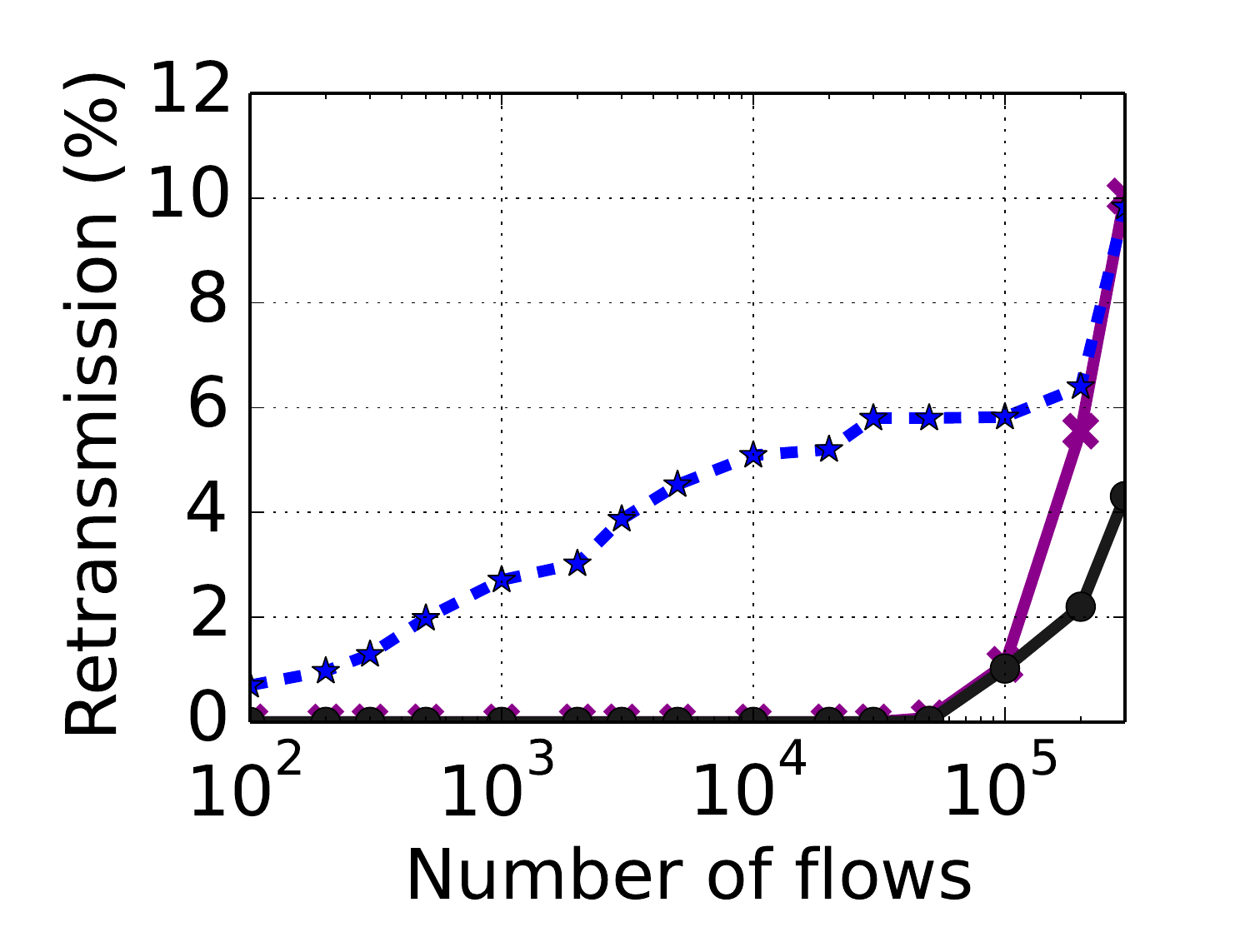}
    \vspace{-0.25in}
    \caption{\small Retransmission}
    \label{fig:summary-retrans}
    \end{subfigure}
\vspace{-0.1in}
\caption{\small Overall performance of the network stack as a function of the number of flows}
\label{fig:summary}
\vspace{-0.3in}
\end{wrapfigure}
\vspace{-0.1in}

We start by measuring the overall performance of the stack with the objective of observing how bottlenecks arise as we increase the number of flows. In particular, we look at aggregate throughput, CPU utilization, average RTT, and retransmissions. Figure~\ref{fig:summary} shows a summary of our results. Our setup can maintain line rate up to around 200k flows (Figure~\ref{fig:summary-throughput}). Thus, we limit our reporting to 300k flows. 

As the number of flows increases, the CPU utilization steadily increases until it becomes the bottleneck. Recall that we are only using a single socket, which means that 50\% utilization means full utilization in our case (Figure~\ref{fig:summary-cpu}). The aggregate throughput shows that the number of bytes per second remains constant. Thus, the increase in CPU utilization is primarily due to the increase in the number of flows handled by the systems.

The most surprising observation is that the average delay introduced by the stack can reach {\emph{one second}} when the stack handles 300k flows, a five orders of magnitude increase from the minimum RTT. There are several problems that can lead to such large delays. The Linux stack is notorious for its inefficiencies due to relying on interrupts, especially on the ingress path \cite{mogul1997eliminating,dpdk,mtcp,brouer2015network}. Further, head-of-line blocking in hardware can add significant delays \cite{stephens2017titan}. Our focus in this paper is to identify problems that are caused by inefficiencies that arise due to the growth in the number of flows. Such problems are likely to occur in the transport and scheduling layers, the layers aware of the number of flows in the system. Our first step is to try to understand which part of the stack is causing these delays, to better understand the impact of the number of flows on the performance of the stack.

Our baseline performance, denoted in the Figure~\ref{fig:summary} by \verb|fq|, is for the case when flows are not rate limited and scheduled following a fair queuing policy, requiring packets to be queued for some flows so that other flows can achieve their fair share. To quantify that delay, we compare the performance of the baseline to a scenario in which each flow is rate limited such that the aggregate rate that is 90\% of NIC capacity, denoted in Figure~\ref{fig:summary} by \verb|per flow rate limit|. Under this scenario, no queuing should happen in the Qdisc as demand is always smaller than the network capacity. Latency drops by an order of magnitude in that scenario at 300k flows and by more at smaller numbers of flows, leading to the conclusion that hundreds of milliseconds of delay are added because of queuing delays at the Qdisc. We further validate this conclusion by employing a Qdisc that implements the CoDel AQM algorithm, configured with a target latency of 100$\mu$s. CoDel drops packets if their queueing delay exceeds the target delay. At 300k flows, the delay of \verb|codel| is lower than the baseline by an order of magnitude, validating our conclusion. Note that CoDel comes at a price of higher CPU utilization due to packet drop and retransmission (Figure~\ref{fig:summary-retrans}). For the rest of the paper, we attempt to better understand the causes of the observed large delays and high CPU utilization at large numbers of flows.

pdf\vspace{-0.1in}
\section{Admission Control to the Stack} \label{sec:admission}

Network stacks are typically optimized to maximize the number of packets per second they can handle, allowing applications unrestricted access to the stack in many cases, especially in Linux. However, as the number of flows increases, applications can overwhelm the stack by generating packets at a larger rate than the network stack can process and transmit them. This congestion, left unchecked, can lead to  hundreds of milliseconds of added delay. Admission control of packets to the stack can avoid this problem by regulating the access of applications to stack resources. Linux already has several such mechanisms, which work well with a relatively small number of flows (e.g., tens of thousands of flows), but fail at large numbers of flows (e.g., hundreds of thousands). We examine admission control mechanisms based on the knob they control. In particular, admission control mechanisms decide three values: 1) the size of each individual packet (the larger the packets the smaller the packet rate for the same byte rate), 2) the total number of admitted packets (i.e., limiting the number of packets through backpressure), and 3) the size of a new batch of admitted packets.

\vspace{-0.1in}
\subsection{Packet Sizing}

The Linux stack implements packet autosizing, an operation that helps improve the pacing function for low throughput flows. Pacing is an integral function for several modern congestion control algorithms including BBR \cite{bbr,fq}. In particular, pacing spreads out packets over time to avoid sending them in bursts. The autosizing algorithm is triggered if a flow is sending at a rate lower than 512 Mbps (i.e., a thousand Maximum Segment Sized (MSS) segments every second, assuming an MSS of 64KB). When triggered, it reduces the size of the segments transmitted every 1ms, where inter-packet gap is enforced through a pacer (e.g., fq \cite{fq}) and packet segmentation to MTU size is done in hardware. Automatic packet sizing can also be beneficial for ensuring fairness between flows \cite{stephens2017titan}.

Autosizing infers the rate of a flow by dividing the number of bytes sent during an RTT (i.e., the \verb|cwnd|) over the measured RTT. This allows for maintaining the same average sending rate while spreading packet transmission over time. The technique provides a tradeoff between CPU utilization and network performance by increasing the number of packets per second handled by the server while lowering the size of bursts the network deals with. The CPU cost of autosizing is affected by the number of flows handled by the server. In particular, the same aggregate rate of 512 Mbps can result in a packet rate of 1k packets per second for one flow or 1M packets per second for 1k flows in the worst case.\footnote{The number of packets is typically much smaller than the worst case scenario due to imperfect pacing. Delays in dispatching packets, resulting from imperfect pacing, require sending larger packets to maintain the correct average rate, leading to a lower packet rate. However, the CPU cost of autosizing increases with the number of flows even with imperfect pacing.} 

\begin{figure}[!t]
\vspace{-0.2in}
\centering
  \begin{subfigure}{0.4\textwidth}
    \centering
    \includegraphics[width=\linewidth,height=35mm]{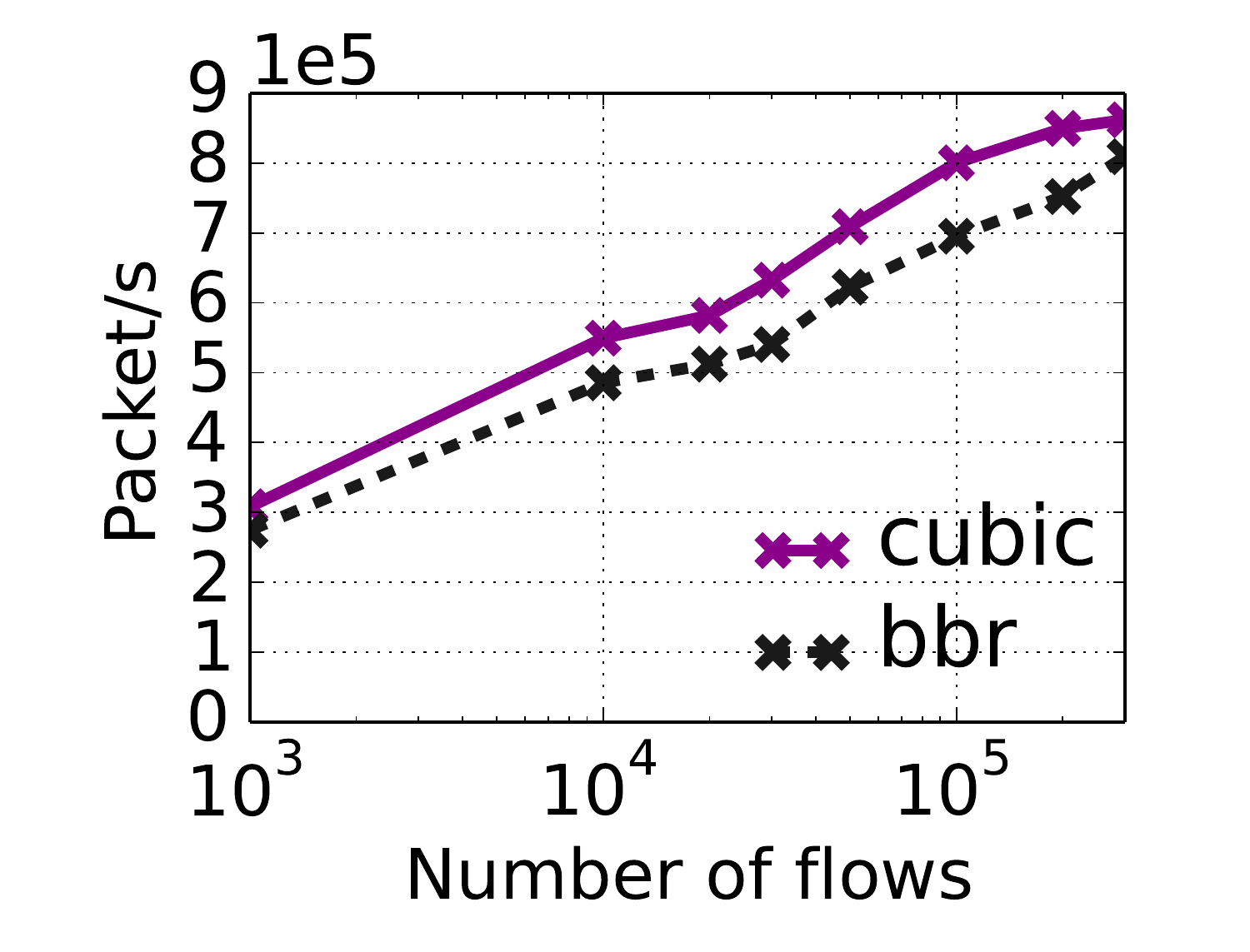}
    \caption{Packet Rate}
    \label{fig:cubic-vs-bbr-loss-pktnum}
    \end{subfigure}
    \begin{subfigure}{0.4\textwidth}
    \centering
    \includegraphics[width=\linewidth,height=35mm]{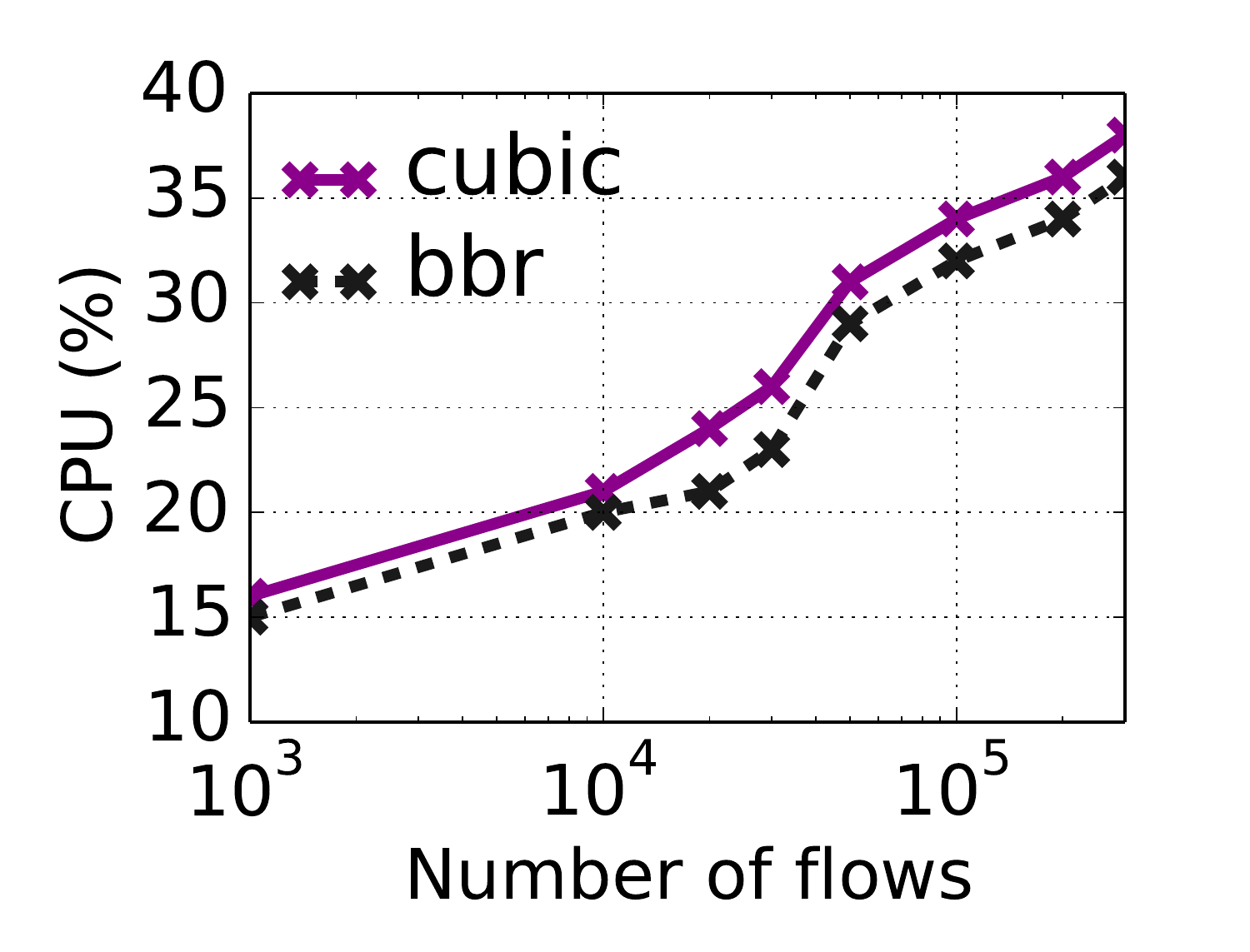}
    \caption{CPU Usage}
    \label{fig:cubic-vs-bbr-loss-cpu}
    \end{subfigure}
\vspace{-0.1in}
\caption{\small CUBIC v.s. BBR with 5\% drop rate. The relationship between number of flows and packet rate is similar at 0\% drop but there is no difference between BBR and CUBIC at 0\% drop rate (Appendix~\ref{app:zero_drop}).}
\label{fig:cubic-vs-bbr-loss}
\vspace{-0.2in}
\end{figure}

This \textbf{overpacing} can overwhelm the stack, leading to an increase in delay (Figure~\ref{fig:summary-rtt}). This leads the autosizing algorithm to misbehave. In particular, the RTT increases when the stack is overloaded, leading to underestimation of the rates of all flows handled by the stack. This causes the autosizing mechanism to reduce the size of bursts unnecessarily, creating more packets, increasing the congestion at the server \cite{zd}. Another side effect of autosizing is causing different congestion control algorithms to have different CPU costs. In particular, algorithms that react more severely to congestion (e.g., CUBIC which halves its window on a packet drop) send at lower rates, forcing autosizing to create more packets. However, algorithms that react mildly to congestion (e.g., BBR), maintain high rates and send lower number of packets. Figure~\ref{fig:cubic-vs-bbr-loss} shows the difference between CUBIC and BBR at 5\% drop rate induced by a \verb|netem| Qdisc at the receiver. We set MTU size to 7000 to eliminate the CPU bottleneck.

Reducing delay introduced in the stack can help autosizing infer the rates of flows more accurately. However, as we will show later, scheduling flows, including delaying packets, is essential to scaling the end host. This means that autosizing-like algorithms need to differentiate between network congestion and end-host congestion. This will be useful in avoiding generating extra packets which might congest the end host but not the network.

\vspace{-0.1in}
\subsection{Backpressure}

When a flow has a packet to send, its thread attempts to enqueue the packet to the packet scheduler (i.e., the Qdisc in the kernel stack). In order to avoid Head-of-Line (HoL) blocking, flows are prevented from sending packets continuously by TCP Small Queue (TSQ). In particular, TSQ limits the number of packets enqueued to the Qdisc to only two packets per flow \cite{tsq}. TSQ offers a rudimentary form of admission control that is based on a per-flow threshold to control the total number of packets in the stack.

\begin{wrapfigure}{r}{0.4\textwidth}
    \centering
    \vspace{-0.5in}
    \includegraphics[width=\linewidth,height=35mm]{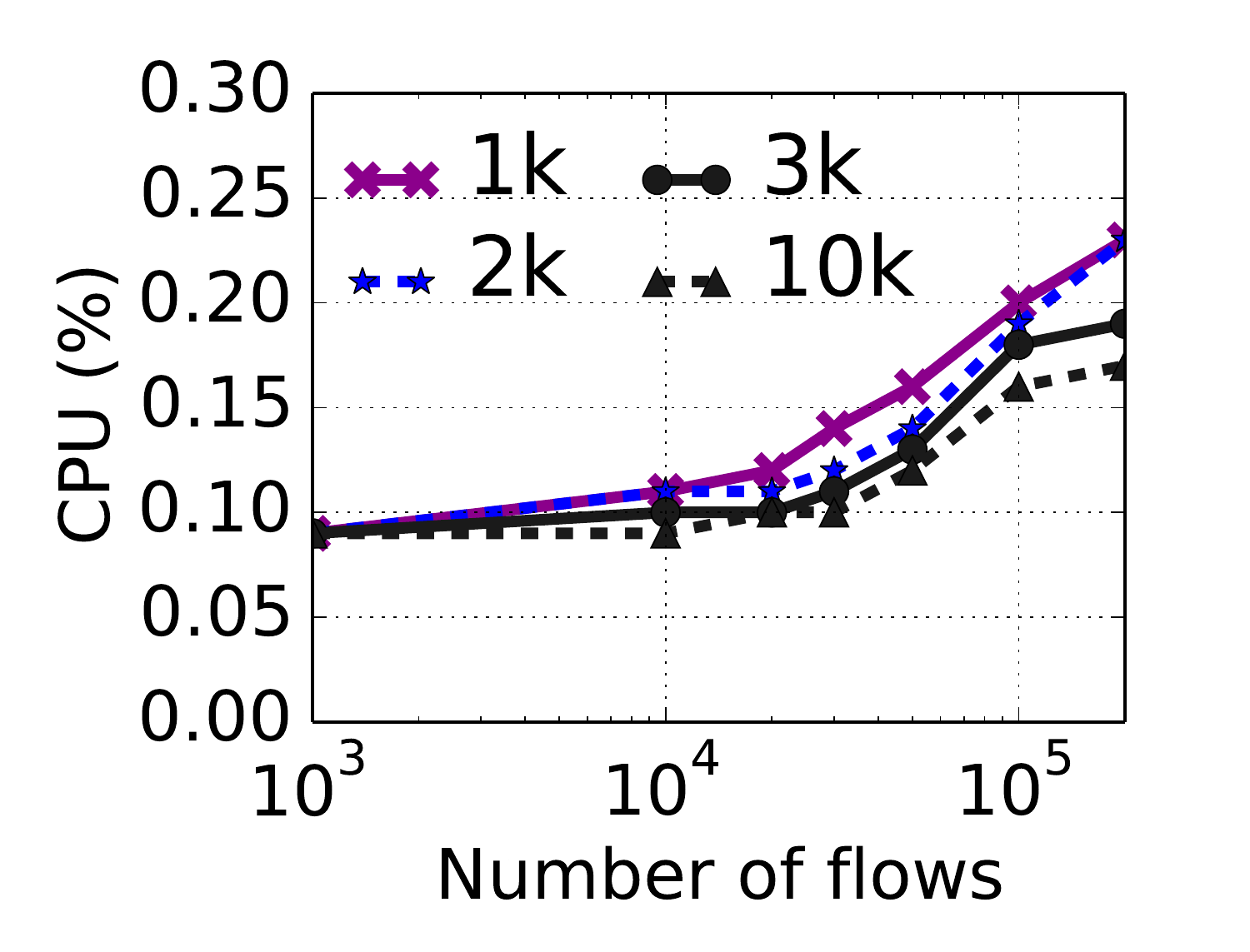}
    \vspace{-0.3in}
    \caption{\small CPU usage as a function of Qdisc queue length}\label{fig:xmit_cpu}
    \vspace{-0.3in}
\end{wrapfigure}
As the number of flows increases, TSQ becomes ineffective because the number of packets admitted to the stack grows with the number of flows. Consequently, the length of the queue in the Qdisc will grow as the number of flows grows, leading to long delays due to bufferbloat. If we limit the queue length of the Qdisc, packets will be dropped at the Qdisc after they are admitted by TSQ. The current approach in Linux is to immediately retry to enqueue the dropped packets, leading to poor CPU utilization as threads keep retrying to enqueue packets. Figure~\ref{fig:xmit_cpu} shows the CPU usage for transmitting packets from the TCP layer to the qdisc with different values of maximum queue length at the qdisc. The CPU usage includes only the operation before enqueuing the packet onto the qdisc. The shorter the queue length, the higher the drop rate, leading to higher CPU utilization.

\begin{wrapfigure}{r}{0.4\textwidth}
\vspace{-0.5in}
\centering
  \begin{subfigure}{0.4\textwidth}
    \centering
    \includegraphics[width=\linewidth, height=35mm]{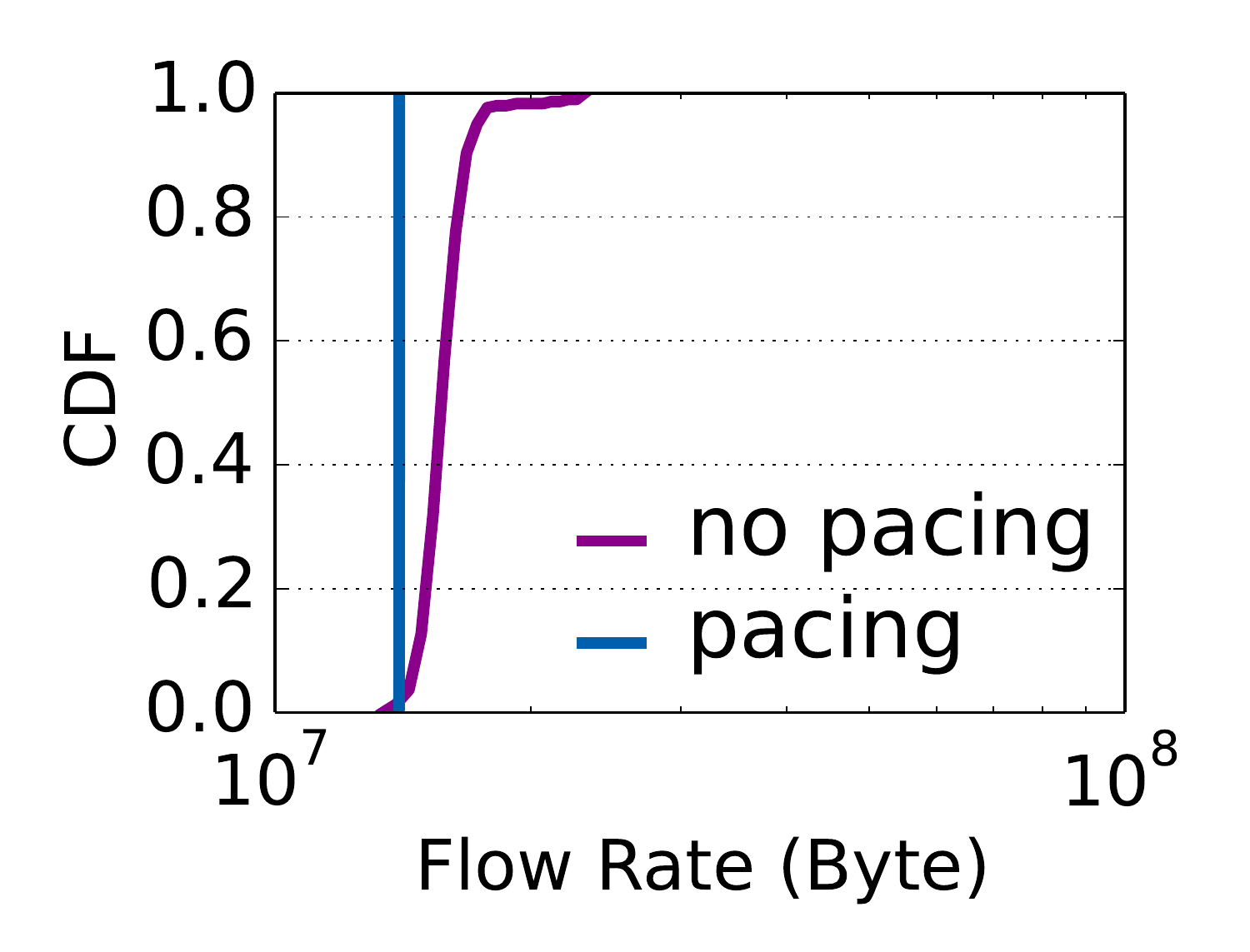}
    \vspace{-0.25in}
    \caption{300 flows}
    \label{fig:300-rate}
    \end{subfigure}
    \begin{subfigure}{0.4\textwidth}
    \centering
    \includegraphics[width=\linewidth,height=35mm]{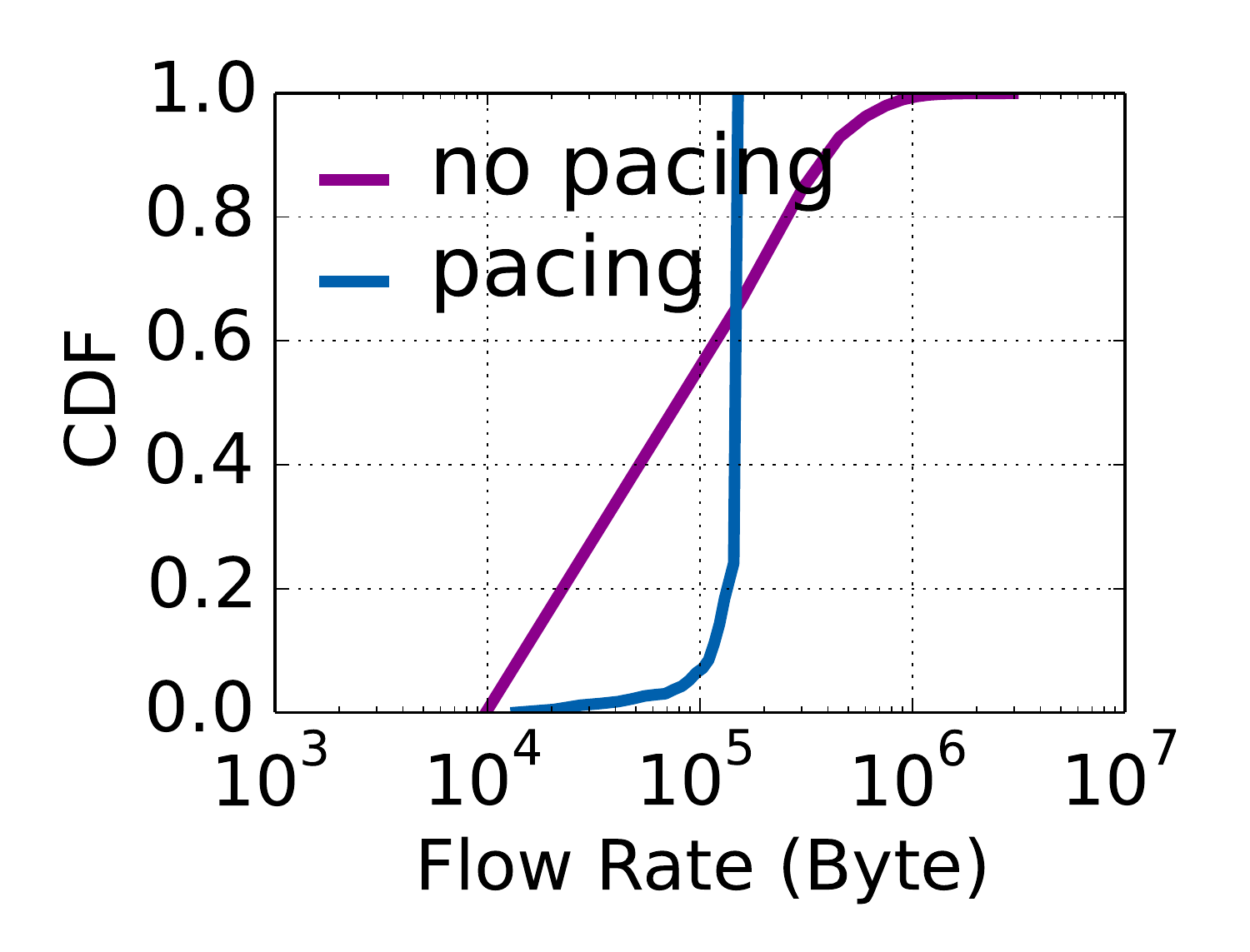}
    \vspace{-0.25in}
    \caption{30k flows}
    \label{fig:30k-rate}
    \end{subfigure}
\vspace{-0.1in}
\caption{\small CDF of flow rate}
\label{fig:unfairness}
\vspace{-0.35in}
\end{wrapfigure}

Another down side of the lack of backpressure is that packet scheduling becomes reliant on thread scheduling. In particular, when a packet is dropped, it is the responsibility of its thread to try to enqueue it again immediately. The frequency at which a thread can ``requeue'' packets depends on the frequency at which it is scheduled. This is problematic because the thread scheduler has no notion of per-flow fairness, leading to severe unfairness between flows. As explained in the previous section, starvation at the Qdisc leads to hundreds of milliseconds of delay on average. We further investigate the effects of this unfairness on per-flow throughput. Figure~\ref{fig:unfairness} compares the CDF of rates achieved when \verb|fq| is used with a small number of 300 and 30k flows. The two scenarios are contrasted with the per-flow pacing scenario which achieves best possible fairness by rate limiting all flows to the same rate, with aggregate rate below NIC capacity, thus avoiding creating a bottleneck at the scheduler. In the 30k flows scenario, the largest rate is two orders of magnitude greater than the smallest rate. This is caused by the batching on the NIC queue. The \verb|net_tx_action| function calls into the Qdisc layer and starts to dequeue skb through the  \verb|dequeue_skb| function. Multiple packets can be returned by some queues, and a list of skb may be sent to NIC, blocking packets from other queues. We observe that there are many more requeue operations in Qdisc when pacing is not used than when pacing is used, indicating that pacing prevents the NIC from being overwhelmed by a subset of queues.

Some previous works address the problem partially by enforcing per-flow scheduling instead of per-packet scheduling and only allowing a flow to enqueue a packet when there is room for it in the scheduler, avoiding unnecessary drops and retries\cite{zd,snap}, however, these works do not consider the interaction between layers that may lead to unfairness when fairness is enforced separately on each layer as we show in this section.

\vspace{-0.17in}
\subsection{Batching Ingress Packets}

\begin{wrapfigure}{r}{0.4\textwidth}
      \centering
      \vspace{-0.5in}
      \includegraphics[width=\linewidth]{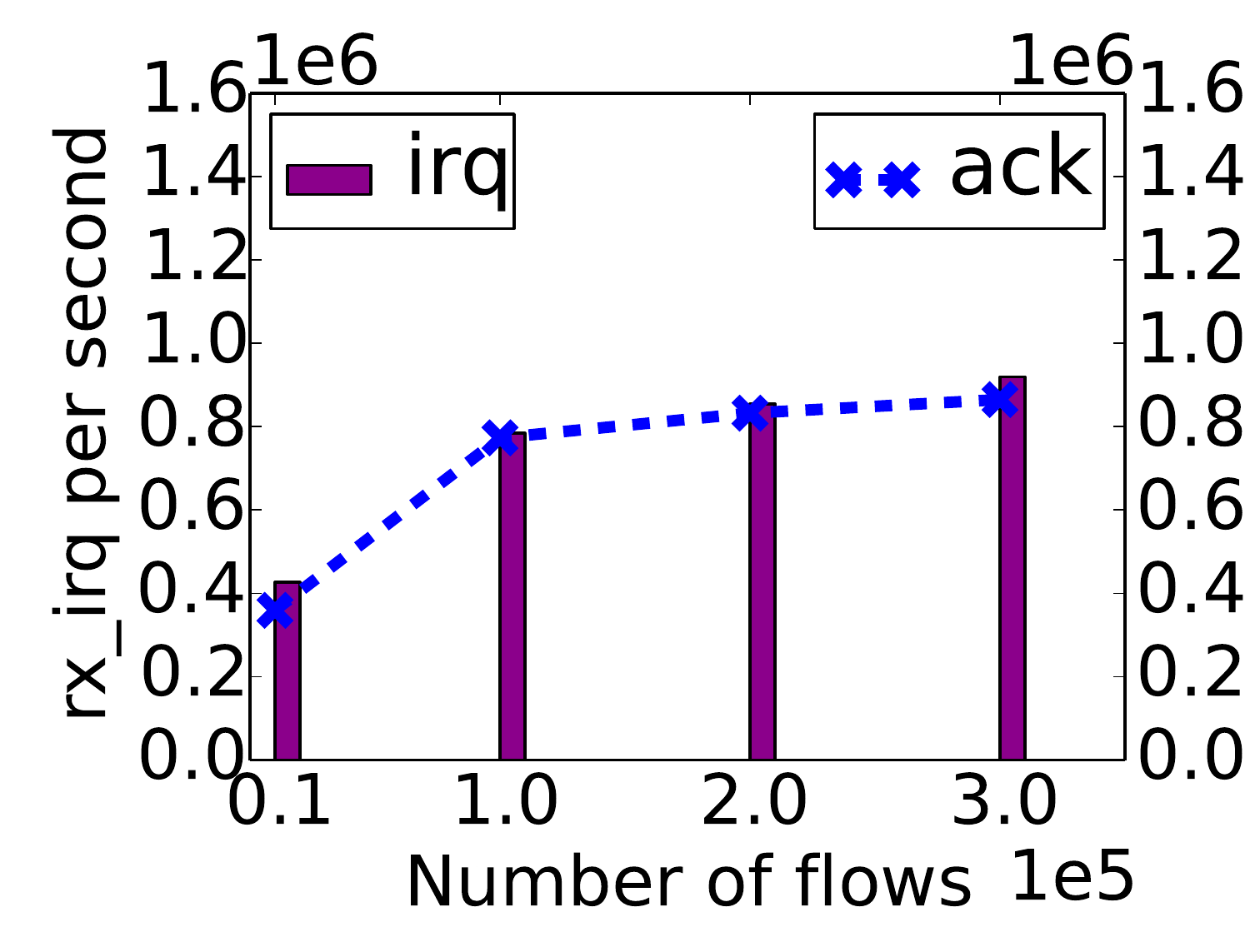}
      \vspace{-0.3in}
      \caption{\small Rates of RX Interrupts and ACKs per second}\label{fig:interrupt-moderation}
      \vspace{-0.3in}
\end{wrapfigure}

The two previous sections discuss controlling the packet rate on the egress path. In this section, we consider controlling the packet rate on the ingress path. It should be noted that although we focus on egress path on server side, ingress path efficiency may also affect the egress path efficiency because delayed ACK caused by CPU saturation can lead to performance degradation in traffic transmission. 

A receiver has little control on the number of incoming packets, aside from flow control. By coalescing packets belonging to the same flow on the ingress path using techniques like LRO, the receiver can improve the CPU efficiency of the receive path by generating less interrupts. 
Batching algorithms deliver packets to the software stack once the number of outstanding packets in the NIC reach a certain maximum batch size or some timer expires. As the number of flows increases, the chances of such coalescing decrease as the likelihood of two incoming packets belong to the same flow decreases (Figure \ref{fig:interrupt-moderation}). In the Linux setting, this is especially bad as increasing the number of incoming packets results in an increase in the number of interrupts, leading to severe degradation in CPU efficiency. 

Better batching techniques that prioritize short flows, and give LRO more time with long flows, can significantly help improve the performance of the ingress path. Some coarse grain adaptive batching techniques have been proposed \cite{sun2014adaptive,li2015adaptive}. However, we believe that better performance can be achieved with fine-grain per-flow adaptive batching, \textbf{requiring coordination between the hardware and software components of the stack}.

pdf

\vspace{-0.1in}

\section{Per-packet Overhead} \label{sec:packet-overhead}

\begin{figure*}[!t]
\vspace{-0.1in}
\begin{minipage}{\textwidth}
\centering
  \begin{subfigure}{0.3\textwidth}
    \centering
    \includegraphics[width=\linewidth]{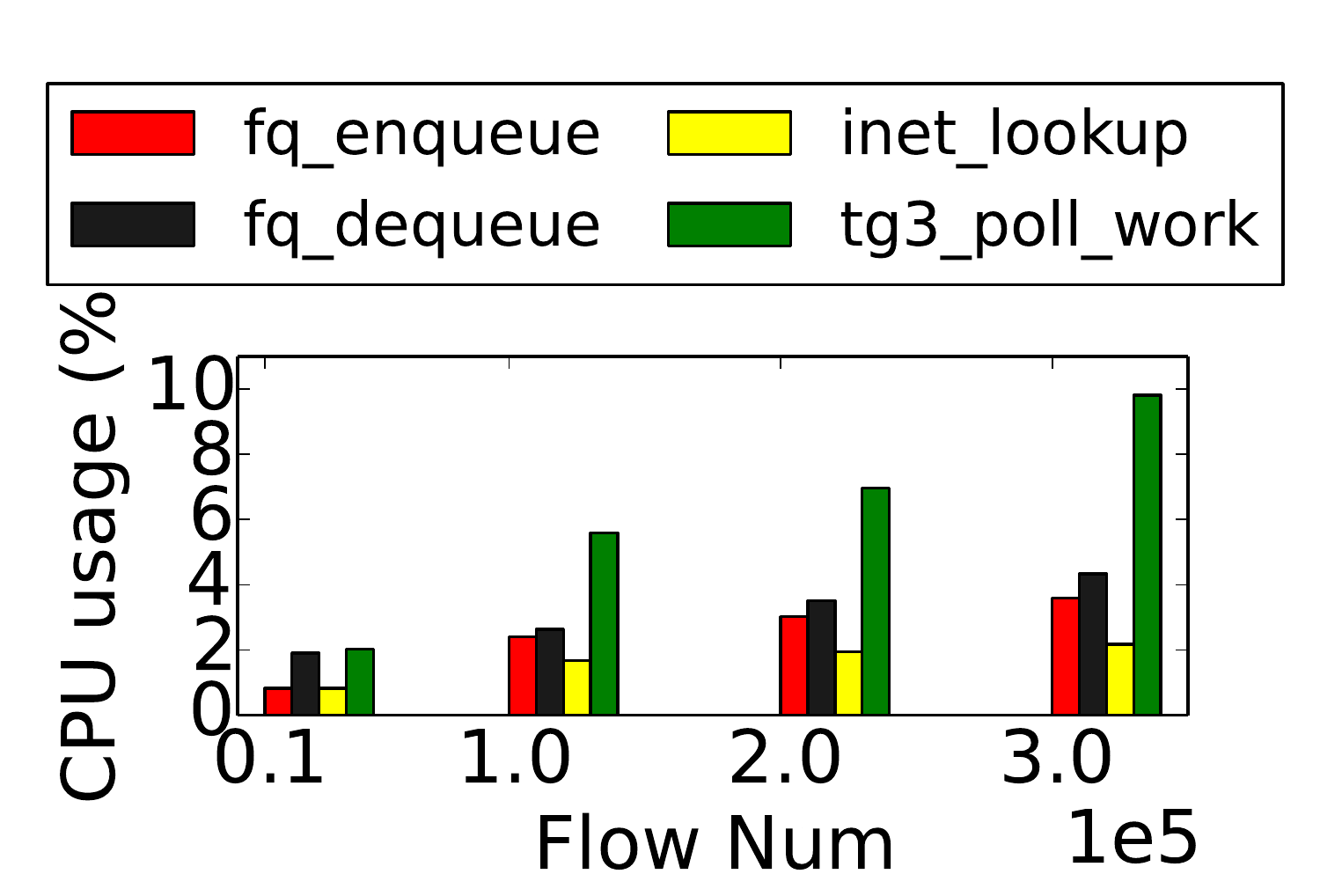}
    \vspace{-0.25in}
    \caption{CPU Usage}
    \label{fig:cpu-break}
    \end{subfigure}
    \begin{subfigure}{0.3\textwidth}
    \centering
    \includegraphics[width=\linewidth]{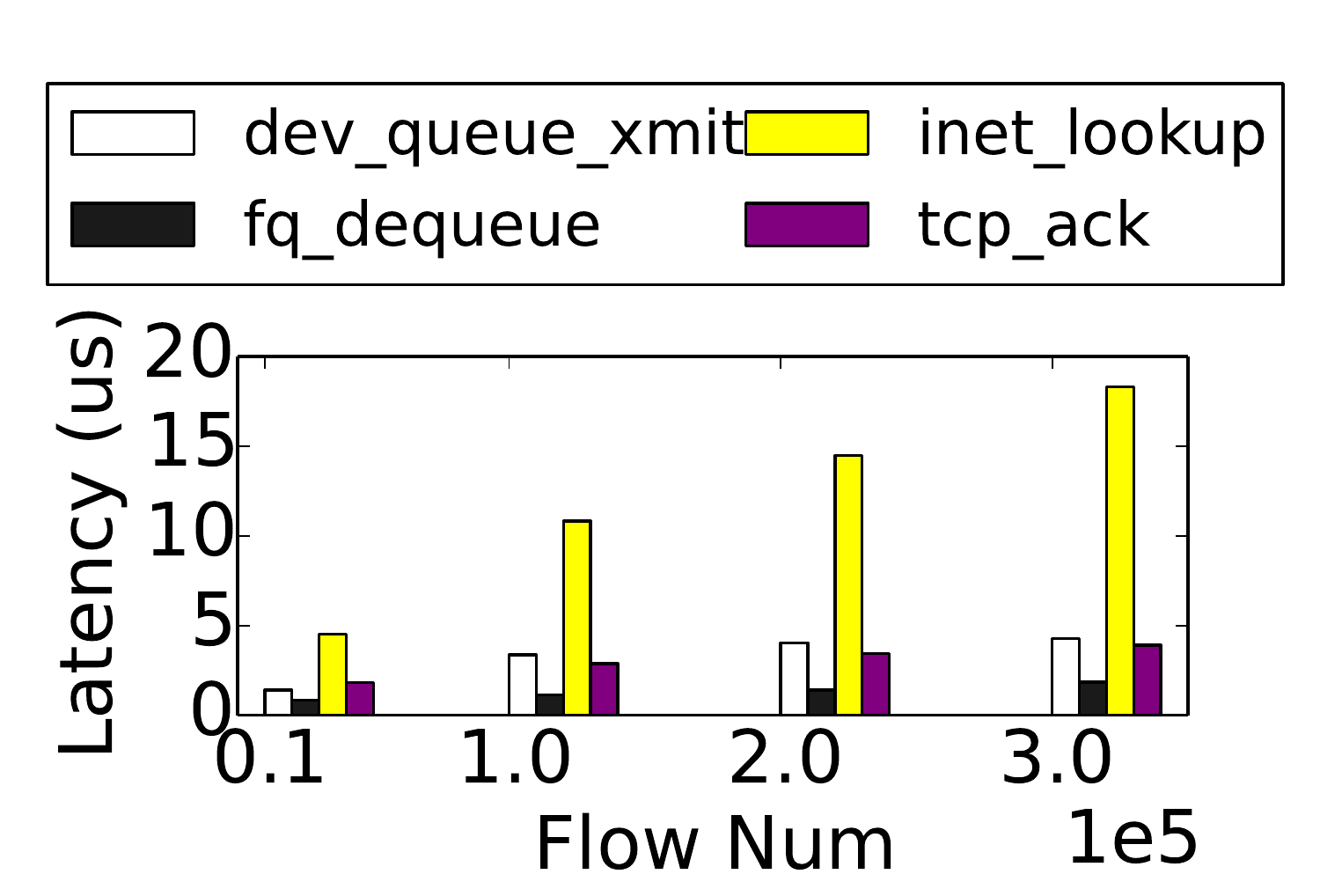}
    \vspace{-0.25in}
    \caption{Function Latency}
    \label{fig:fun-latency}
    \end{subfigure}
    \begin{subfigure}{0.3\textwidth}
    \centering
    \includegraphics[width=\linewidth]{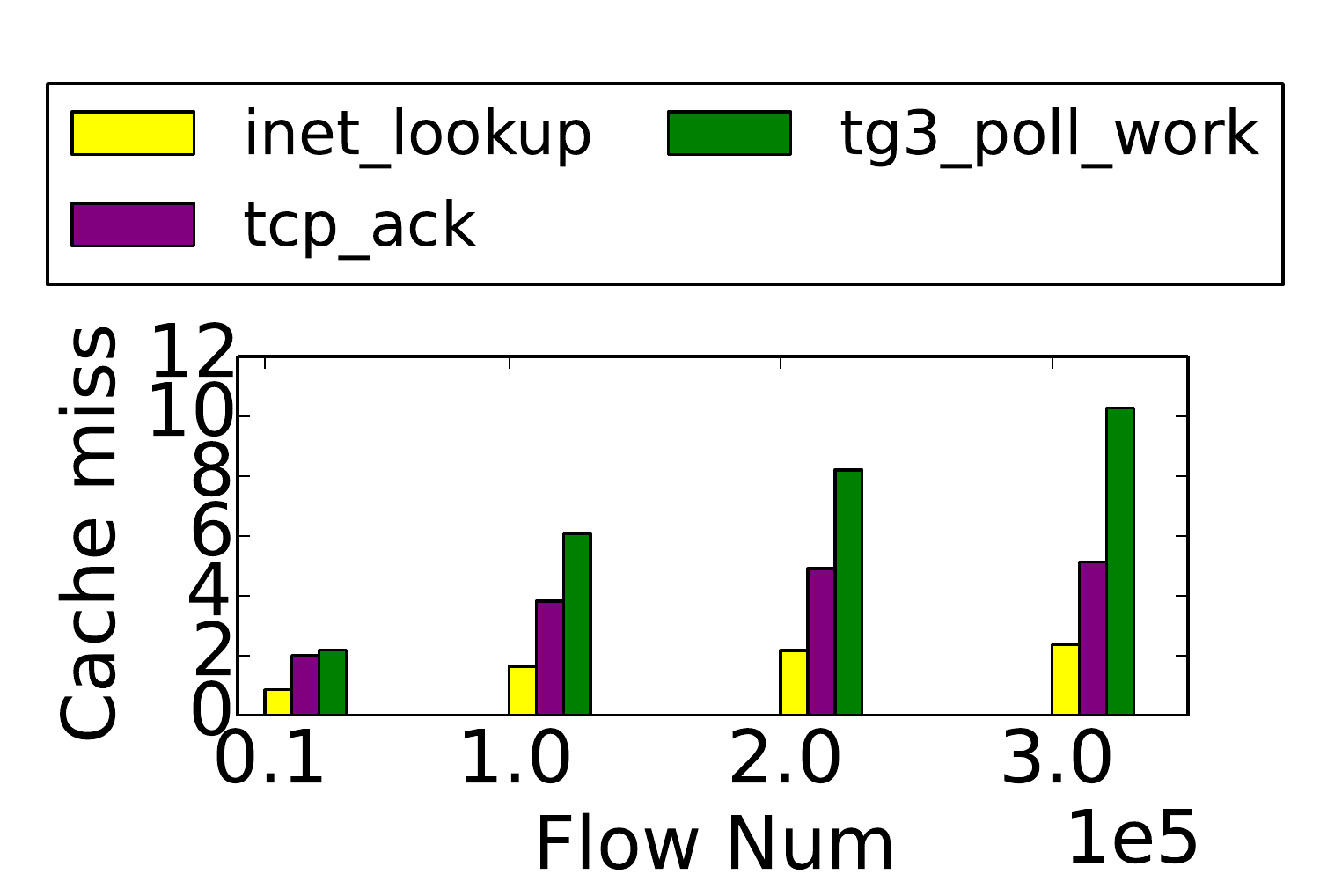}
    \vspace{-0.25in}
    \caption{Cache Misses}
    \label{fig:fun-cache}
    \end{subfigure}

\vspace{-0.1in}
\caption{Function profiling}
\label{fig:function-break}
\end{minipage}
\vspace{-0.1in}
\end{figure*}

To identify the operations whose overhead increases as the number of flows increases, we use \verb|perf|~\cite{perf-cmd} and observe the CPU utilization and latency of different kernel functions as we change the number of flows. The CPU utilization results show the aggregated CPU usage by all flows. We keep the aggregate data rate the same and only change the number of flows. Our goal is to find the operations whose computational complexity is a function of the number of flows. Operations that are bottlenecked on a different type of resource will have higher latency as we increase the number of flows. Figures~\ref{fig:cpu-break} and \ref{fig:fun-latency} show the top four functions in each category. There is an overlap between functions with high latency and functions with high CPU utilization; this is typical because high CPU utilization can lead to high latency (e.g., \verb|fq_dequeue| and \verb|inet_lookup|). However, there are functions with high latency but low CPU utilization (e.g., \verb|tcp_ack| and \verb|dev_queue_xmit|). Through further profiling of the code of these functions, we find that there are two types of bottlenecks that arise: cache pressure and lock contention. Note that the overhead of the \verb|tg3_poll_work| function is part of inefficiency of the Linux reception path \cite{benvenuti2006understanding} and is not the focus of our work.

\noindent\textbf{Data structures:} There are two operations whose complexity is a function of the number of flows: packet scheduling and packet demultiplexing. The overhead of packet scheduling is captured by the CPU utilization of \verb|fq_enqueue| and \verb|fq_dequeue|. The two functions handle adding and removing packets to the \verb|fq| Qdisc, which sorts flows in a red-black tree based on the soonest transmission time of their packets. The overhead of enqueue and dequeue operations in $O(\log(n))$, where $n$ is the number of flows. The overhead of packet demultiplexing is captured by the CPU utilization of \verb|inet_lookup| which matches incoming packets to their flows using a hashmap. 
In the case of collision, finding a match requires processing information of flows whose hash collide. This increases the cache miss ratio of the function (Figure~\ref{fig:fun-cache}), further increasing the latency of the function.

\begin{wrapfigure}{r}{0.38\textwidth}
    \vspace{-0.4in}
    \centering
    \includegraphics[width=\linewidth]{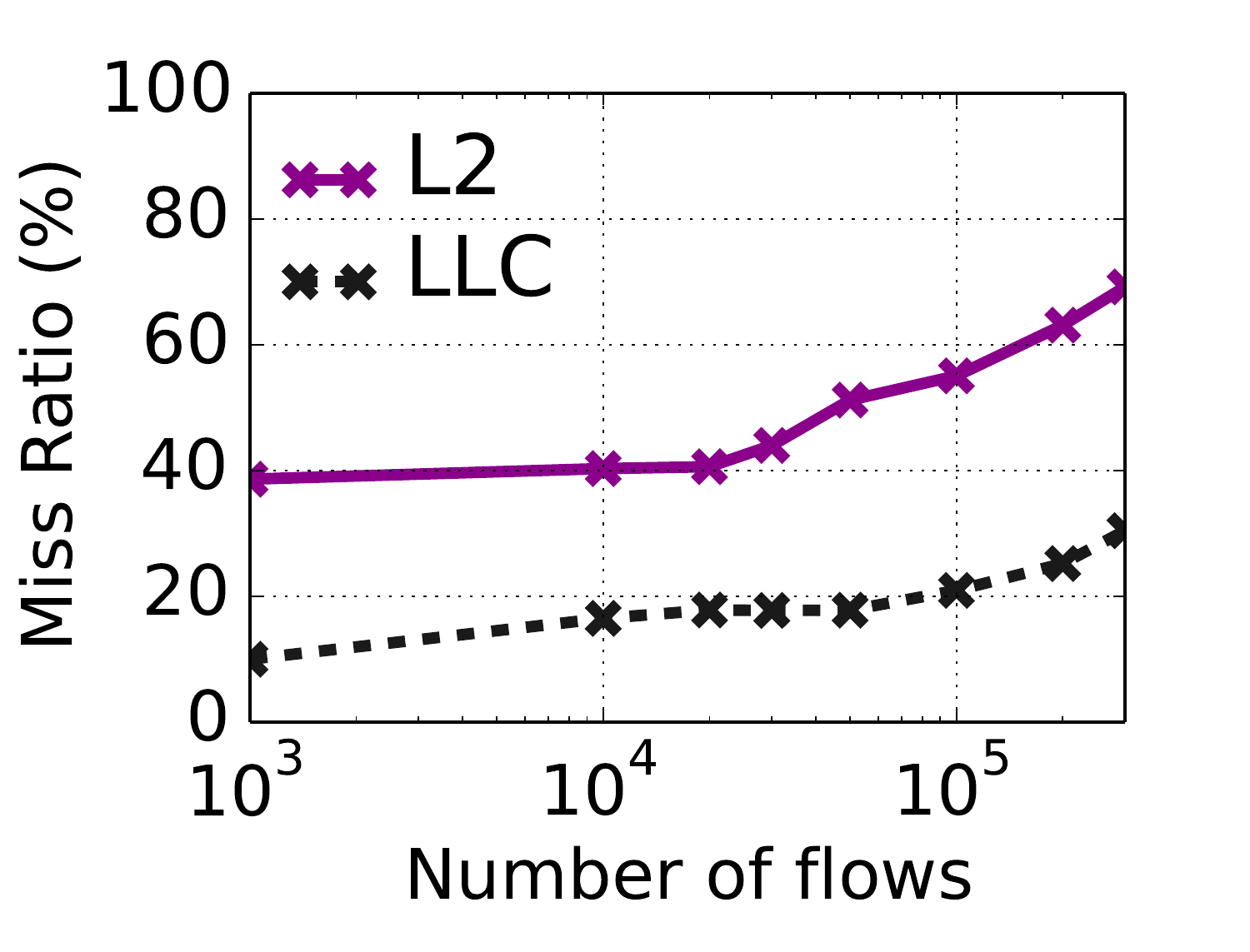}
    \vspace{-0.3in}
    \caption{\small Aggregate cache misses}\label{fig:cache-misses}
    \vspace{-0.3in}
\end{wrapfigure}  

Some approximation scheduling algorithms have been proposed to reduce the data structure overhead \cite{qfq,carousel,eiffel}, but their main focus is to improve FQ. Data structure overhead requires reexamining all complex data structures used in the stack, taking into account that the stack can process millions of packets per second coming from millions of flows.

\noindent\textbf{Cache pressure:} One of the functions with the highest cache miss ratio is \verb|tcp_ack|, which clears the TCP window based on received acknowledgements. The function does not use any complex data structures or wait on locks so the high cache miss stems from the overhead of fetching flow information and modifying it. 
As shown in Figure \ref{fig:cache-misses}, the cache miss ratio in both L2 cache and Last Level Cache (LLC) increases as the number of flows increases. While cache misses are not a huge bottleneck in our setting, we believe that as the number of flows increases, with tighter requirements on latency, cache miss ratio will have to be minimized.

\begin{wrapfigure}{r}{0.4\textwidth}
    \vspace{-0.3in}
    \centering
    \includegraphics[width=\linewidth]{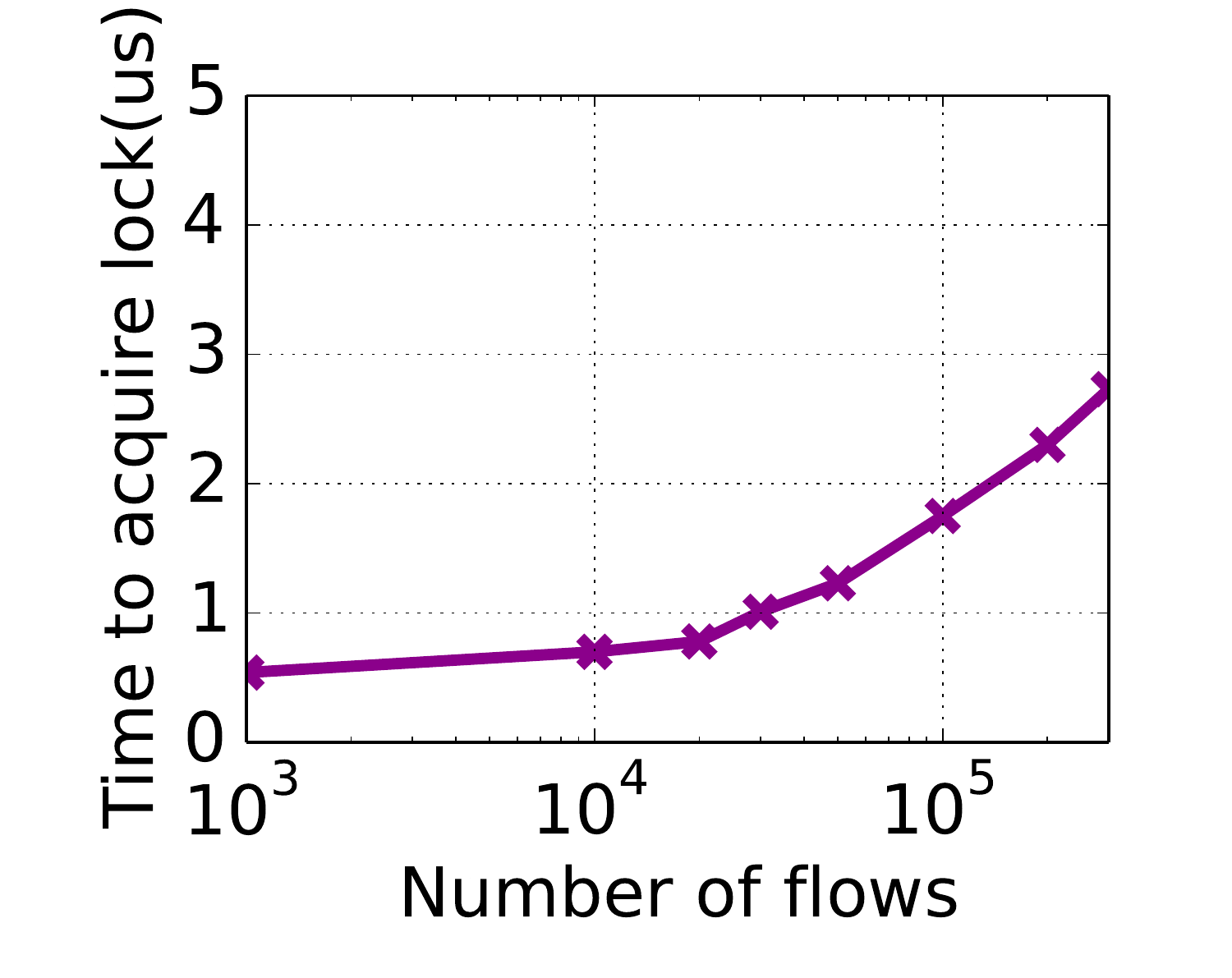}
    \vspace{-0.3in}
    \caption{\small Time to acquire qdisc lock}\label{fig:lock-time}
    \vspace{-0.3in}
\end{wrapfigure}    

\noindent\textbf{Lock contention:} Another source of increased latency is lock contention when accessing shared resources. Our experiment confirms that the biggest critical section in the networking stack is the one used to protect access to the qdisc, done in \verb|dev_queue_xmit|. The overhead of acquiring the qdisc lock is well documented \cite{radhakrishnan2014senic,carousel}, and increasing the number of flows exacerbates the problem, even with constant packet rate. Figure \ref{fig:lock-time} shows that as the time to acquire lock increases by 4 times as the number of flow increases from 1k to 300k. Another factor contributing to the increase in lock acquisition time is the increase in packet rate which we have shown to increase as the number of flows increases (Figure~\ref{fig:cubic-vs-bbr-loss-pktnum}). Distributed and lazy coordination between independent queues can help alleviate the problem by reducing the need for locking \cite{carousel,multi_queue}.

\vspace{-0.15in}

\section{Related Work}
As we present throughout the paper, there has been significant work improving different components of the stack including scheduling \cite{qfq,carousel,eiffel,multi_queue} and backpressure \cite{zd}. However, they fail to consider the interactions between different components, and none of the existing optimized components was tested with a load larger than 50k flows. Our work defines a broader category of limitations and looks at the complicated interaction between different components.

Much of the focus of the previous work has been on scaling servers in terms of aggregate traffic intensity in terms of packets transmitted per second, while maintaining low latency \cite{dpdk,netmap,flexnic,ix,shenango}. Some recent proposals address scaling the whole stack to handle a large number of flows \cite{mtcp,rotaru2017reliable,tas,acceltcp}. mTcp \cite{mtcp} is a scalable user-space TCP/IP stack built over kernel-bypass packet I/O engines, but the evaluation was only performed at a maximum of 16k flows. Further, it focuses on improving connection locality and reducing system overhead without paying much attention to scheduling and backpressure. Other systems are evaluated at a few thousands flows \cite{tas} and up to twenty thousand flows \cite{rotaru2017reliable,acceltcp,stackmap,netmap}. These systems improve specific functionality (e.g., RPC performance or transport layer performance) by dedicating network
interfaces to individual application or by optimizing the kernel TCP/IP stack, with typical emphasis on short lived flows. In this paper, we are more concerned with scaling to hundreds of thousands of long-lived flows where transport and scheduling are implemented. To the best of our knowledge, this is the first such study. 

Another observation is that hardware offload solutions \cite{pieo,loom,firestone} alone cannot completely solve the problem. Careful hardware design can help reduce the latency of complex operations \cite{pieo}. However, data structure issues do not disappear when implemented in hardware. In addition, admission control requires careful coordination between the the software part of the stack, including the application, and the hardware part of the stack.

\vspace{-0.1in}
\section{Relevance of Findings to Other Stacks}
\label{sec:generalize}

In this paper, we focus on the Linux stack because of its ubiquitous usage in both industry and academia. However, most of our findings focus on abstract functions that are needed in a stack in order to efficiently handle a large number of flows. For example, admission control can avoid overwhelming the stack resources by relying on per-flow scheduling and accurate batching sizing. The lack of similar functions in any stack can lead to performance degradation as the number of flows grows. Further, the need for better data structures for scheduling and demultiplexing can lead to significant CPU savings. Contrarily, some of the problems we define are Linux specific, arising from components developed by companies to handle their specific workloads. For example, autosizing was developed by Google, making problems like overpacing a Linux-specific problem.

Some stacks inherently solve some of the problems we have identified. For instance, Snap \cite{snap} provides per-flow scheduling providing efficient backpressure. Further, stacks that rely on lightweight threading and asynchronous messages like Snap and Shenango might not suffer significant performance degradation due to lock contention. However, none of them handles all problems
The goal of our work is to identify abstract functions that stacks will have to implement in order to scale.

Some of the problems we have identified are only exposed at a very large number of flows. To the best of our knowledge, these problems are yet to be handled by any stack. For instance, delays introduced due to cache misses will require innovation in speculative pre-fetching based on network behavior. Further, network accelerators and programmable hardware components will require new techniques to coordinate their behavior with changes in the load generated by the software component of the stack. 

\vspace{-0.1in}
\section{Conclusion}
In this paper, we identify the different bottlenecks that arise when we scale the number of flows to hundreds of thousands in a fully implemented stack. 
As we present throughout the paper, there have been efforts to address some of the individual problems in isolation. However, integrating and testing such solutions at the scale of hundreds of thousands to millions of long-lived simultaneously-active flows remains an open problem. 
We hope that this paper sheds some light on the pain points that stack designers should pay attention to when building next generation stacks that scale to terabits per second and millions of flows.


\bibliographystyle{splncs04}
\bibliography{stack_measurements}

\begin{thebibliography}{10}
\providecommand{\url}[1]{\texttt{#1}}
\providecommand{\urlprefix}{URL }
\providecommand{\doi}[1]{https://doi.org/#1}

\bibitem{100gbps_nic}
High-performance, feature-rich netxtreme® e-series dual-port 100g pcie
  ethernet nic,
  \url{https://www.broadcom.com/products/ethernet-connectivity/network-adapters/100gb-nic-ocp/p2100g}

\bibitem{dpdk}
{Intel DPDK: Data plane development kit}. \url{https://www.dpdk.org/} (2014)

\bibitem{8207825}
Ieee standard for ethernet - amendment 10: Media access control parameters,
  physical layers, and management parameters for 200 gb/s and 400 gb/s
  operation. IEEE Std 802.3bs-2017 (Amendment to IEEE 802.3-2015 as amended by
  IEEE's 802.3bw-2015, 802.3by-2016, 802.3bq-2016, 802.3bp-2016, 802.3br-2016,
  802.3bn-2016, 802.3bz-2016, 802.3bu-2016, 802.3bv-2017, and IEEE
  802.3-2015/Cor1-2017) pp. 1--372 (2017)

\bibitem{core_trend}
Microprocessor trend data (2018),
  \url{https://github.com/karlrupp/microprocessor-trend-data}

\bibitem{ethernet_roadmap}
IEEE 802.3 Industry Connections Ethernet Bandwidth Assessment Part II  (2020)

\bibitem{dstat-cmd}
{dstat-Linux man page}. \url{https://linux.die.net/man/1/dstat} (2020)

\bibitem{fq_codel}
{FlowQueue-Codel}.
  \url{https://tools.ietf.org/id/draft-ietf-aqm-fq-codel-02.html} (2020)

\bibitem{neper}
{neper: a Linux networking performance tool}.
  \url{https://github.com/google/neper} (2020)

\bibitem{netflix_bitrate}
{Netflix Help Center: Internet Connection Speed Recommendations} (2020),
  \url={https://help.netflix.com/en/node/306}

\bibitem{netstat-cmd}
{netstat-Linux man page}. \url{https://linux.die.net/man/8/netstat} (2020)

\bibitem{perf-cmd}
{Perf Manual}. \url{https://www.man7.org/linux/man-pages/man1/perf.1.html}
  (2020)

\bibitem{ss-cmd}
{ss-Linux man page}. \url{https://linux.die.net/man/8/ss} (2020)

\bibitem{ix}
Belay, A., Prekas, G., Klimovic, A., Grossman, S., Kozyrakis, C., Bugnion, E.:
  $\{$IX$\}$: A protected dataplane operating system for high throughput and
  low latency. In: 11th $\{$USENIX$\}$ Symposium on Operating Systems Design
  and Implementation ($\{$OSDI$\}$ 14). pp. 49--65 (2014)

\bibitem{benvenuti2006understanding}
Benvenuti, C.: Understanding Linux network internals. " O'Reilly Media, Inc."
  (2006)

\bibitem{brouer2015network}
Brouer, J.D.: Network stack challenges at increasing speeds. In: Proc. Linux
  Conf. pp. 12--16 (2015)

\bibitem{bbr}
Cardwell, N., Cheng, Y., Gunn, C.S., Yeganeh, S.H., Jacobson, V.: Bbr:
  Congestion-based congestion control. Queue  \textbf{14}(5),  20--53 (2016)

\bibitem{cavalcanti2009optimizing}
Cavalcanti, F.R.P., Andersson, S.: Optimizing wireless communication systems,
  vol.~386. Springer (2009)

\bibitem{qfq}
{Checconi}, F., {Rizzo}, L., {Valente}, P.: Qfq: Efficient packet scheduling
  with tight guarantees. IEEE/ACM Transactions on Networking  \textbf{21}(3)
  (2013)

\bibitem{chen2011peer}
Chen, Q.C., Yang, X.H., Wang, X.L.: A peer-to-peer based passive web crawling
  system. In: 2011 International Conference on Machine Learning and
  Cybernetics. vol.~4, pp. 1878--1883. IEEE (2011)

\bibitem{tsq}
Dumazet, E., Corbet, J.: {TCP small queues}.
  \url{https://lwn.net/Articles/507065/} (2012)

\bibitem{fq}
Dumazet, E., Corbet, J.: Tso sizing and the fq scheduler.
  \url{https://lwn.net/Articles/564978/} (2013)

\bibitem{firestone}
Firestone, D., Putnam, A., Mundkur, S., Chiou, D., Dabagh, A., Andrewartha, M.,
  Angepat, H., Bhanu, V., Caulfield, A., Chung, E., et~al.: Azure accelerated
  networking: Smartnics in the public cloud. In: 15th $\{$USENIX$\}$ Symposium
  on Networked Systems Design and Implementation ($\{$NSDI$\}$ 18). pp. 51--66
  (2018)

\bibitem{geer2005chip}
Geer, D.: Chip makers turn to multicore processors. IEEE Computer  \textbf{38}
  (2005)

\bibitem{multi_queue}
Hedayati, M., Shen, K., Scott, M.L., Marty, M.: Multi-queue fair queuing. In:
  2019 {USENIX} Annual Technical Conference ({USENIX} {ATC} 19) (2019)

\bibitem{hock2019tcp}
Hock, M., Veit, M., Neumeister, F., Bless, R., Zitterbart, M.: Tcp at 100
  gbit/s--tuning, limitations, congestion control. In: 2019 IEEE 44th
  Conference on Local Computer Networks (LCN). pp.~1--9. IEEE (2019)

\bibitem{mtcp}
Jeong, E., Wood, S., Jamshed, M., Jeong, H., Ihm, S., Han, D., Park, K.: mtcp:
  a highly scalable user-level $\{$TCP$\}$ stack for multicore systems. In:
  11th $\{$USENIX$\}$ Symposium on Networked Systems Design and Implementation
  ($\{$NSDI$\}$ 14). pp. 489--502 (2014)

\bibitem{kalia2019datacenter}
Kalia, A., Kaminsky, M., Andersen, D.: Datacenter rpcs can be general and fast.
  In: 16th $\{$USENIX$\}$ Symposium on Networked Systems Design and
  Implementation ($\{$NSDI$\}$ 19). pp. 1--16 (2019)

\bibitem{flexnic}
Kaufmann, A., Peter, S., Sharma, N.K., Anderson, T., Krishnamurthy, A.: High
  performance packet processing with flexnic. In: Proceedings of the
  Twenty-First International Conference on Architectural Support for
  Programming Languages and Operating Systems. pp. 67--81 (2016)

\bibitem{tas}
Kaufmann, A., Stamler, T., Peter, S., Sharma, N.K., Krishnamurthy, A.,
  Anderson, T.: Tas: Tcp acceleration as an os service. In: Proceedings of the
  Fourteenth EuroSys Conference 2019. pp. 1--16 (2019)

\bibitem{li2015adaptive}
Li, Y., Cornett, L., Deval, M., Vasudevan, A., Sarangam, P.: Adaptive interrupt
  moderation (Apr~14 2015), uS Patent 9,009,367

\bibitem{snap}
Marty, M., de~Kruijf, M., Adriaens, J., Alfeld, C., Bauer, S., Contavalli, C.,
  Dalton, M., Dukkipati, N., Evans, W.C., Gribble, S., Kidd, N., Kononov, R.,
  Kumar, G., Mauer, C., Musick, E., Olson, L., Rubow, E., Ryan, M., Springborn,
  K., Turner, P., Valancius, V., Wang, X., Vahdat, A.: Snap: A microkernel
  approach to host networking. In: Proceedings of the 27th ACM Symposium on
  Operating Systems Principles. p. 399–413. SOSP ’19 (2019)

\bibitem{mogul1997eliminating}
Mogul, J.C., Ramakrishnan, K.: Eliminating receive livelock in an
  interrupt-driven kernel. ACM Transactions on Computer Systems
  \textbf{15}(3),  217--252 (1997)

\bibitem{acceltcp}
Moon, Y., Lee, S., Jamshed, M.A., Park, K.: Acceltcp: Accelerating network
  applications with stateful {TCP} offloading. In: 17th {USENIX} Symposium on
  Networked Systems Design and Implementation ({NSDI} 20). pp. 77--92 (2020)

\bibitem{shenango}
Ousterhout, A., Fried, J., Behrens, J., Belay, A., Balakrishnan, H.: Shenango:
  Achieving high {CPU} efficiency for latency-sensitive datacenter workloads.
  In: Proc. of {USENIX} {NSDI} '19 (2019)

\bibitem{radhakrishnan2014senic}
Radhakrishnan, S., Geng, Y., Jeyakumar, V., Kabbani, A., Porter, G., Vahdat,
  A.: $\{$SENIC$\}$: Scalable $\{$NIC$\}$ for end-host rate limiting. In: 11th
  $\{$USENIX$\}$ Symposium on Networked Systems Design and Implementation
  ($\{$NSDI$\}$ 14). pp. 475--488 (2014)

\bibitem{netmap}
Rizzo, L.: Netmap: a novel framework for fast packet i/o. In: 21st USENIX
  Security Symposium (USENIX Security 12). pp. 101--112 (2012)

\bibitem{rotaru2017reliable}
Rotaru, M., Olariu, F., Onica, E., Rivi{\`e}re, E.: Reliable messaging to
  millions of users with migratorydata. In: Proceedings of the 18th
  ACM/IFIP/USENIX Middleware Conference: Industrial Track. pp.~1--7 (2017)

\bibitem{carousel}
Saeed, A., Dukkipati, N., Valancius, V., Lam, T., Contavalli, C., Vahdat, A.:
  {Carousel: Scalable Traffic Shaping at End-Hosts}. In: Proc. of ACM SIGCOMM
  '17 (2017)

\bibitem{eiffel}
Saeed, A., Zhao, Y., Dukkipati, N., Zegura, E., Ammar, M., Harras, K., Vahdat,
  A.: Eiffel: Efficient and flexible software packet scheduling. In: Proc. of
  {USENIX} {NSDI} '19 (2019)

\bibitem{pieo}
Shrivastav, V.: Fast, scalable, and programmable packet scheduler in hardware.
  In: Proceedings of the ACM Special Interest Group on Data Communication.
  SIGCOMM ’19 (2019)

\bibitem{loom}
Stephens, B., Akella, A., Swift, M.: Loom: Flexible and efficient $\{$NIC$\}$
  packet scheduling. In: 16th $\{$USENIX$\}$ Symposium on Networked Systems
  Design and Implementation ($\{$NSDI$\}$ 19). pp. 33--46 (2019)

\bibitem{stephens2017titan}
Stephens, B., Singhvi, A., Akella, A., Swift, M.: Titan: Fair packet scheduling
  for commodity multiqueue nics. In: 2017 $\{$USENIX$\}$ Annual Technical
  Conference (USENIX ATC '17). pp. 431--444 (2017)

\bibitem{sun2014adaptive}
Sun, L., Kostic, P.: Adaptive hardware interrupt moderation (Jan~2 2014), uS
  Patent App. 13/534,607

\bibitem{stackmap}
Yasukata, K., Honda, M., Santry, D., Eggert, L.: Stackmap: Low-latency
  networking with the $\{$OS$\}$ stack and dedicated nics. In: 2016
  $\{$USENIX$\}$ Annual Technical Conference ($\{$USENIX$\}$$\{$ATC$\}$ 16).
  pp. 43--56 (2016)

\bibitem{zhang2017tuning}
Zhang, T., Wang, J., Huang, J., Chen, J., Pan, Y., Min, G.: Tuning the
  aggressive tcp behavior for highly concurrent http connections in
  intra-datacenter. IEEE/ACM Transactions on Networking  \textbf{25}(6),
  3808--3822 (2017)

\bibitem{zd}
Zhao, Y., Saeed, A., Zegura, E.W., Ammar, M.H.: {zD: A Scalable Zero-Drop
  Network Stack at End Hosts}. In: Proceedings of the 15th International
  Conference on Emerging Networking Experiments And Technologies (CoNEXT). pp.
  220--232. CoNEXT '19, {ACM} (2019). \doi{10.1145/3359989.3365425}

\end{thebibliography}
\appendix

\section{Linux Stack Overview}

\begin{figure}[!t]
\centering
\includegraphics[width=0.5\linewidth]{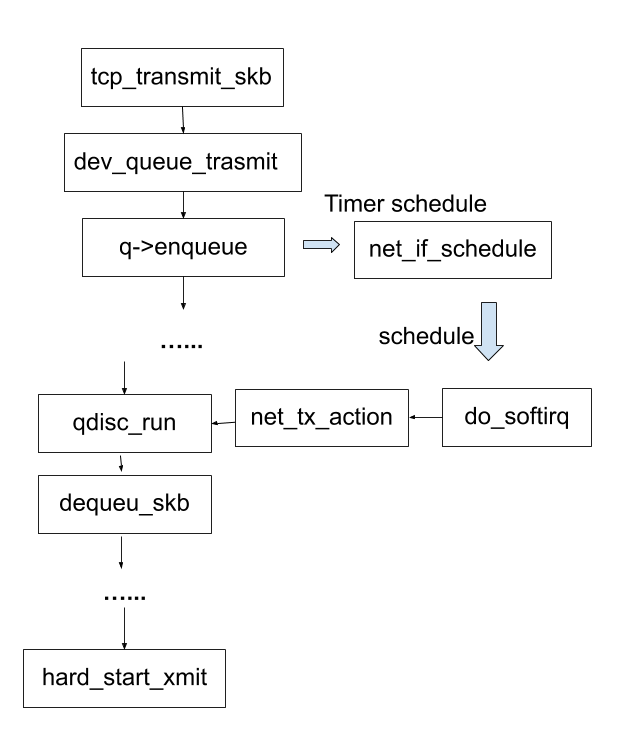}
\vspace{-0.1in}
\caption{Packet Transmission}
\label{fig:pkt-transmit}
\vspace{-0.1in}
\end{figure}

Packet transmission in an end-host refers to the process of a packet traversing from user space, to kernel space, and finally to NIC in packet transmission process. The application generates a packet and copies it into the kernel space TCP buffer. Packets from the TCP buffer are then queued into Qdisc. Then there are two ways to a dequeue packet from the Qdisc to the driver buffer: 1)dequeue a packet immediately, and 2) schedule a packet to be dequeued later through softriq, which calls net\_tx\_action to retrieve packet from qdisc (Figure \ref{fig:pkt-transmit}).

\begin{figure}[!t]
\begin{minipage}{0.48\textwidth}
  \begin{subfigure}{0.48\textwidth}
    \centering
    \includegraphics[width=\linewidth]{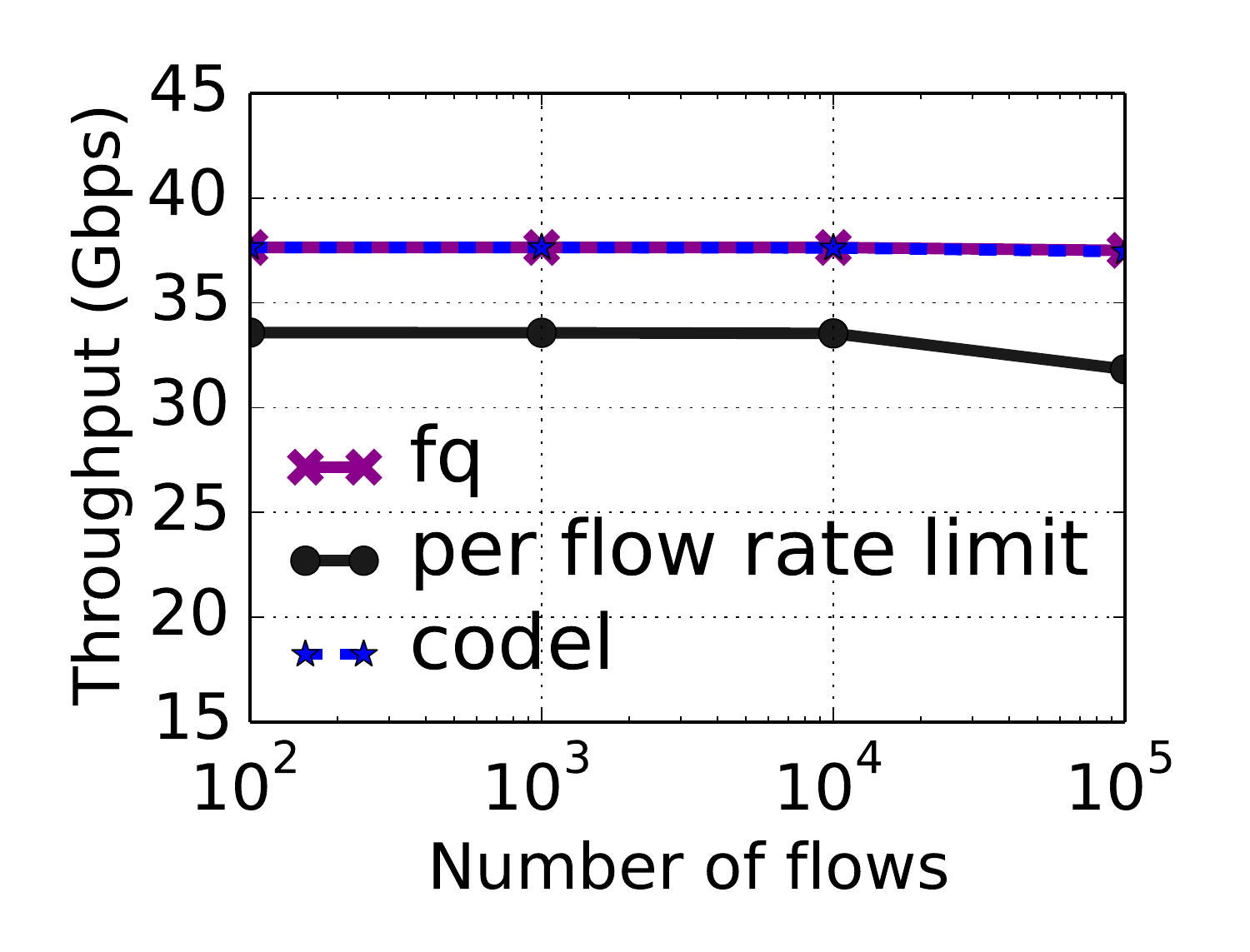}
    \vspace{-0.25in}
    \caption{Throughput}
    \label{fig:summary-throughput-1500}
    \end{subfigure}
    \begin{subfigure}{0.48\textwidth}
    \centering
    \includegraphics[width=\linewidth]{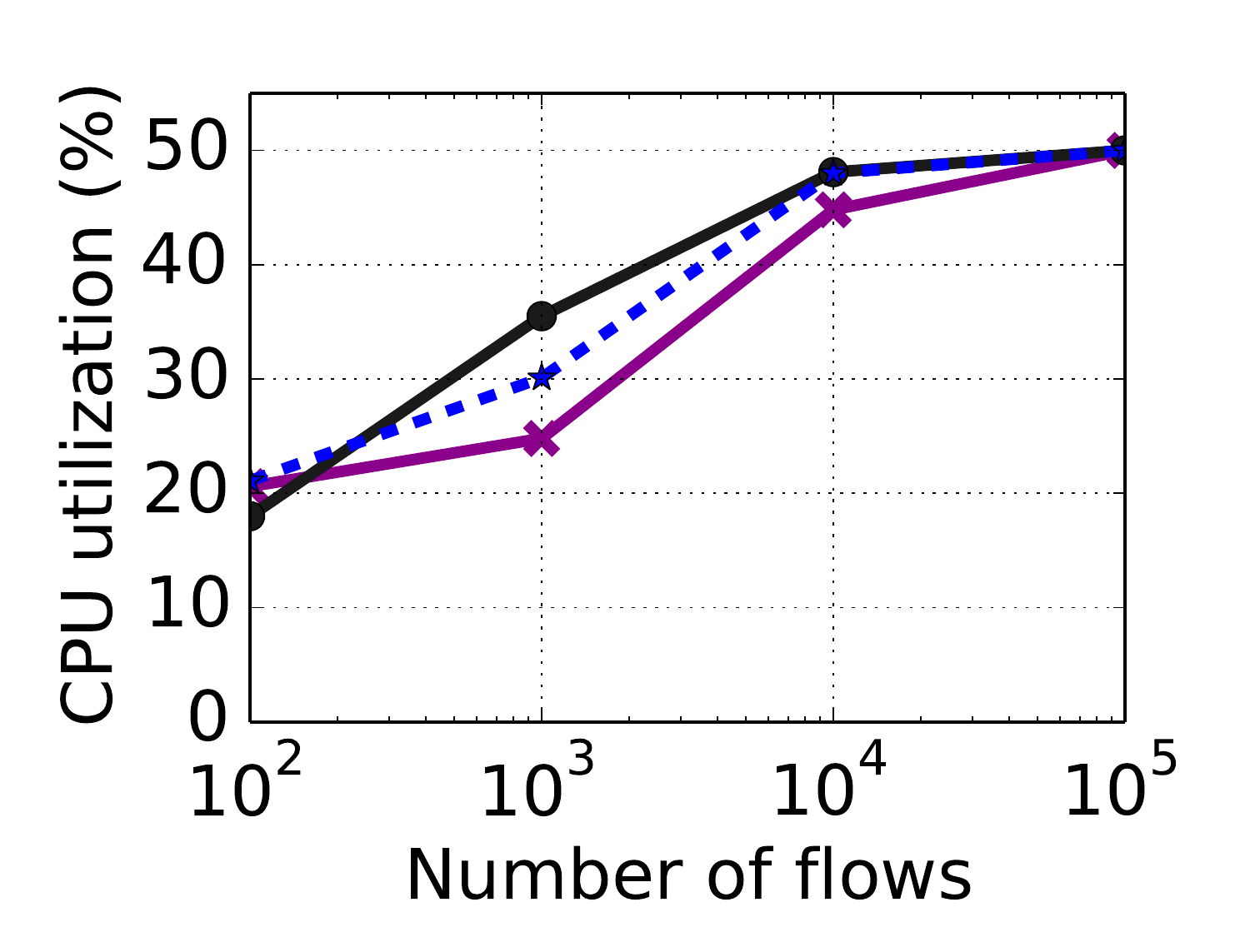}
    \vspace{-0.25in}
    \caption{CPU Usage}
    \label{fig:summary-cpu-1500}
    \end{subfigure}
    \begin{subfigure}{0.48\textwidth}
    \centering
    \includegraphics[width=\linewidth]{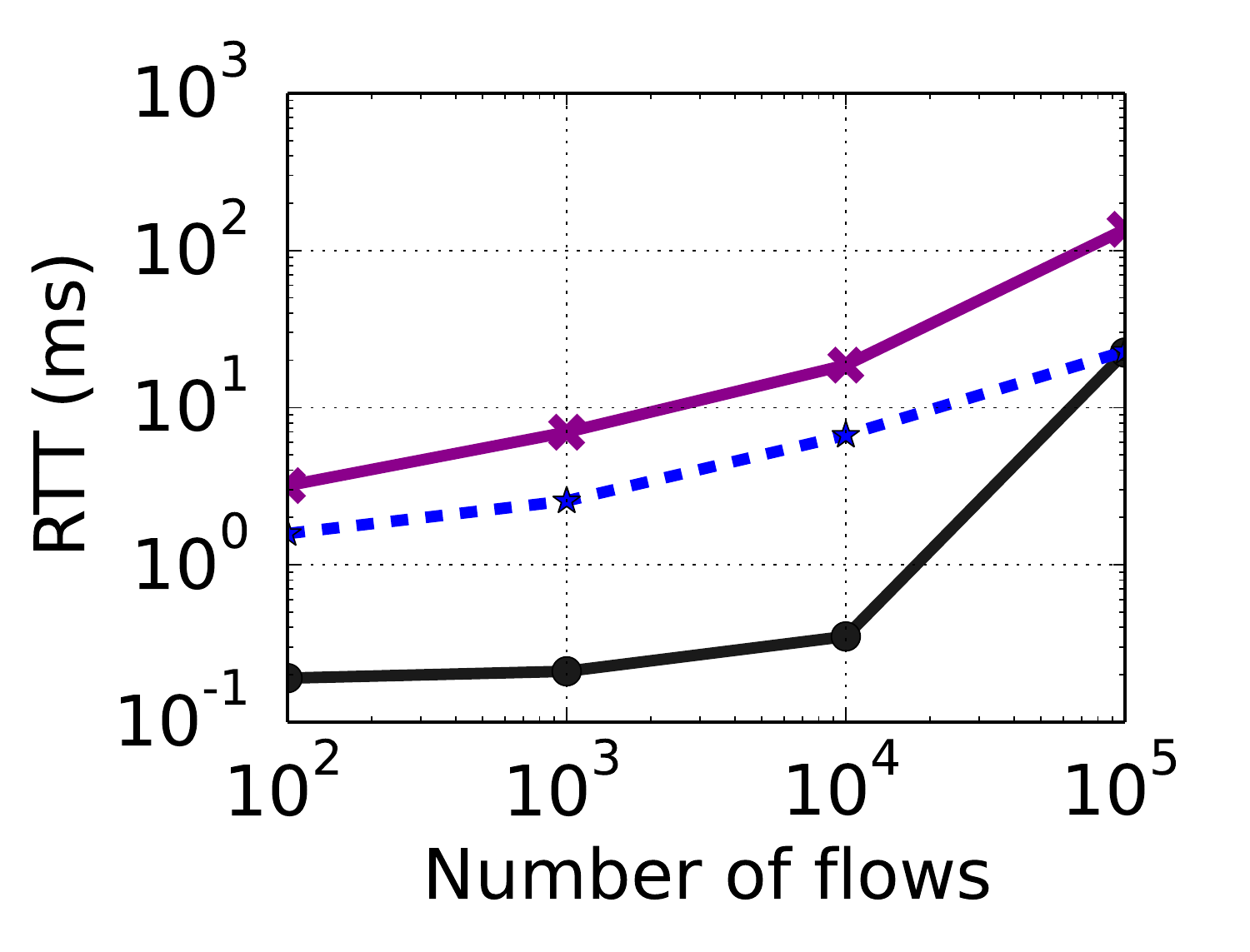}
    \vspace{-0.25in}
    \caption{RTT}
    \label{fig:summary-rtt-1500}
    \end{subfigure}
    \begin{subfigure}{0.48\textwidth}
    \centering
    \includegraphics[width=\linewidth]{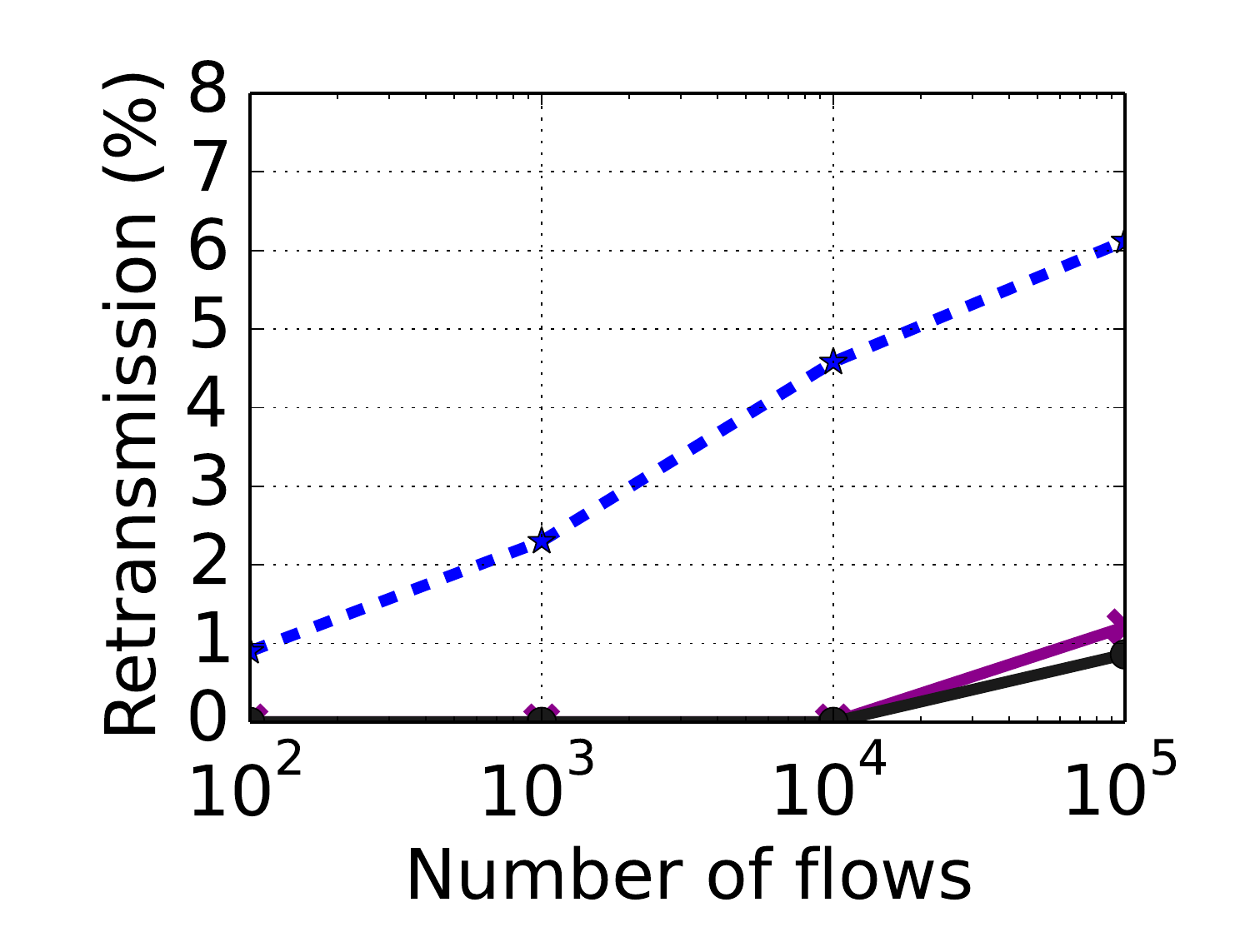}
    \vspace{-0.25in}
    \caption{Retransmission}
    \label{fig:summary-retrans-1500}
    \end{subfigure}
\vspace{-0.1in}
\caption{Overall performance of the network stack as a function of the number of flows with fixed TSO disabled and 1500 MTU size}
\label{fig:summary-1500}
\end{minipage}
\hfill
\begin{minipage}{0.48\textwidth}
  \begin{subfigure}{0.48\textwidth}
    \centering
    \includegraphics[width=\linewidth]{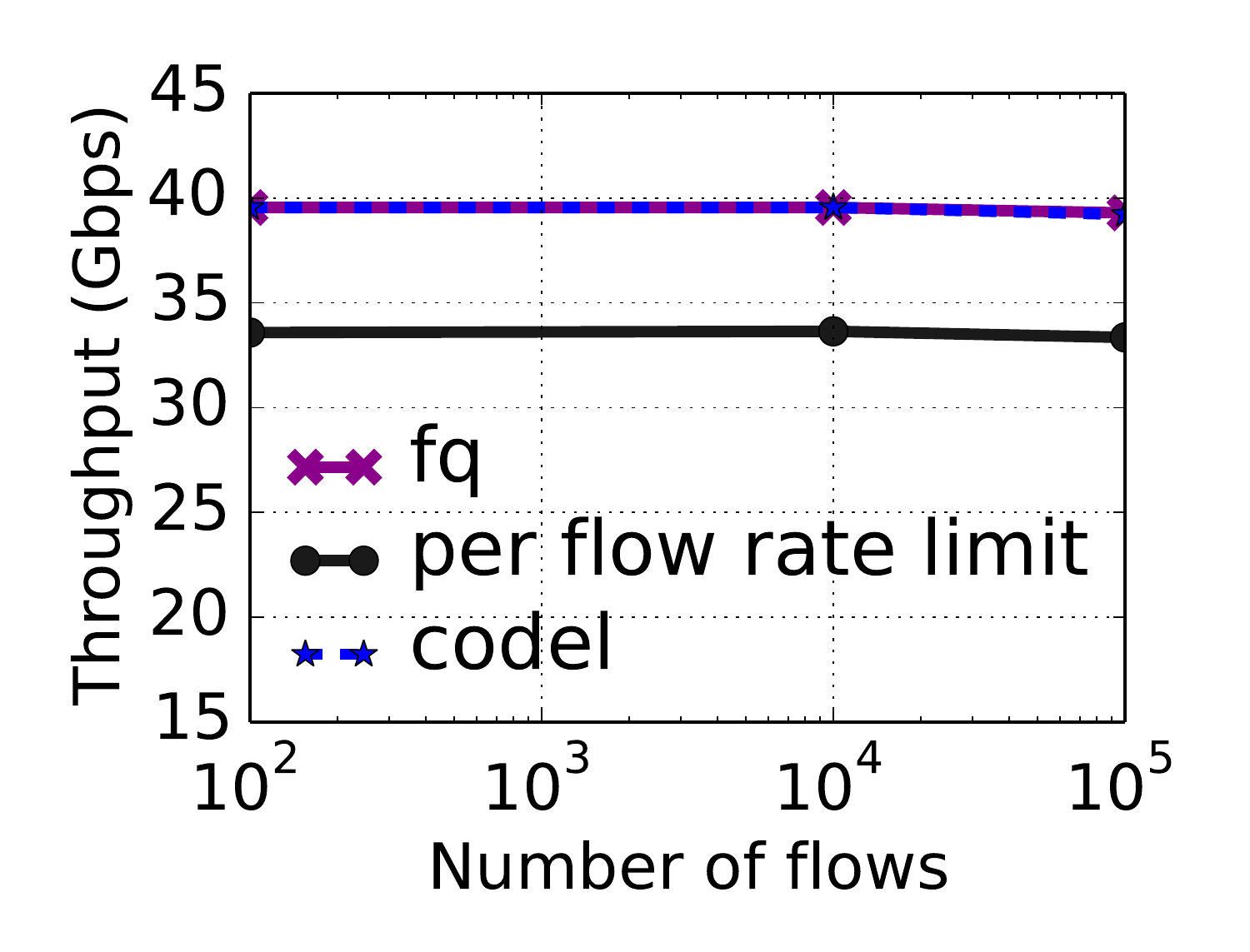}
    \vspace{-0.25in}
    \caption{Throughput}
    \label{fig:summary-throughput-9000}
    \end{subfigure}
    \begin{subfigure}{0.48\textwidth}
    \centering
    \includegraphics[width=\linewidth]{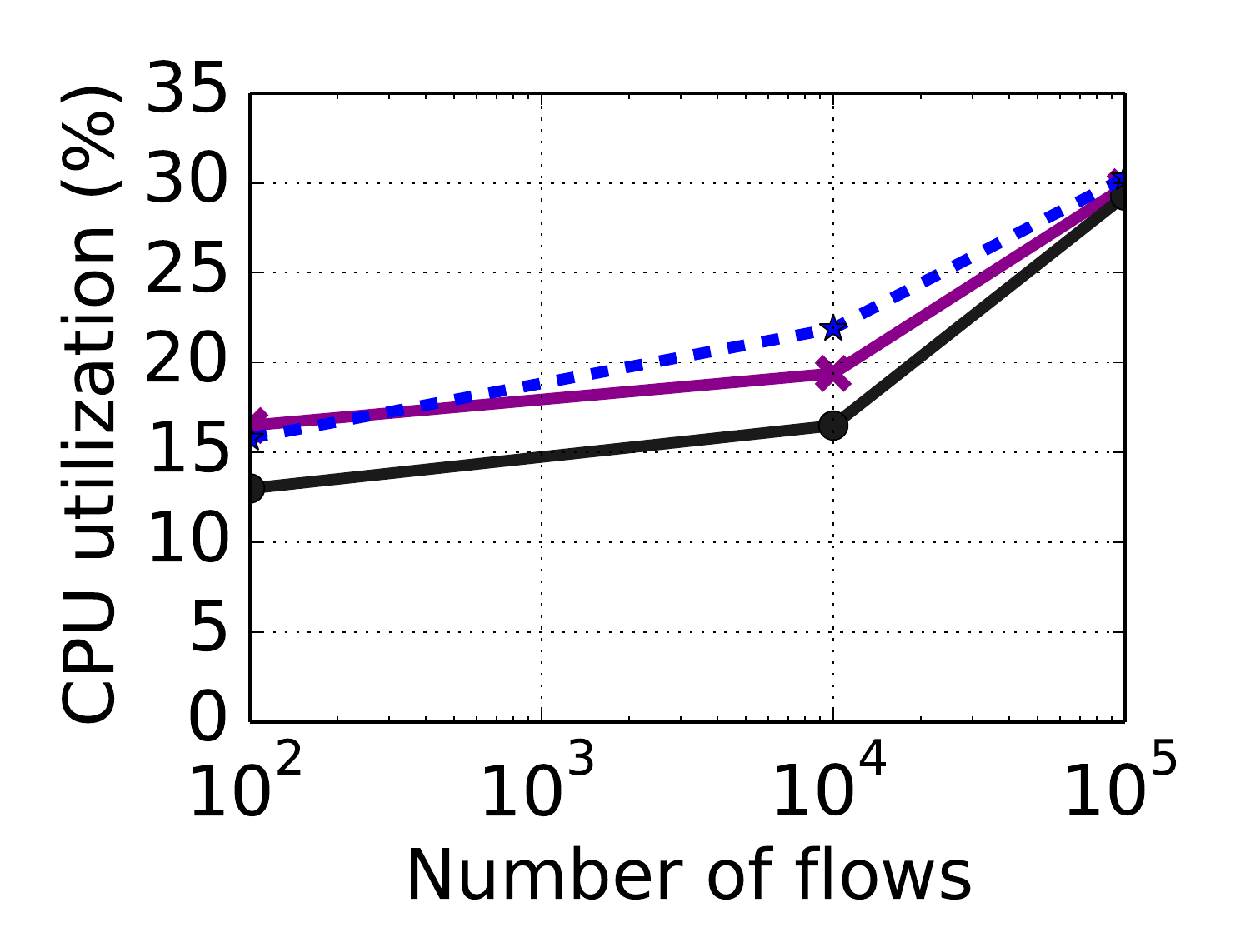}
    \vspace{-0.25in}
    \caption{CPU Usage}
    \label{fig:summary-cpu-9000}
    \end{subfigure}
    \begin{subfigure}{0.48\textwidth}
    \centering
    \includegraphics[width=\linewidth]{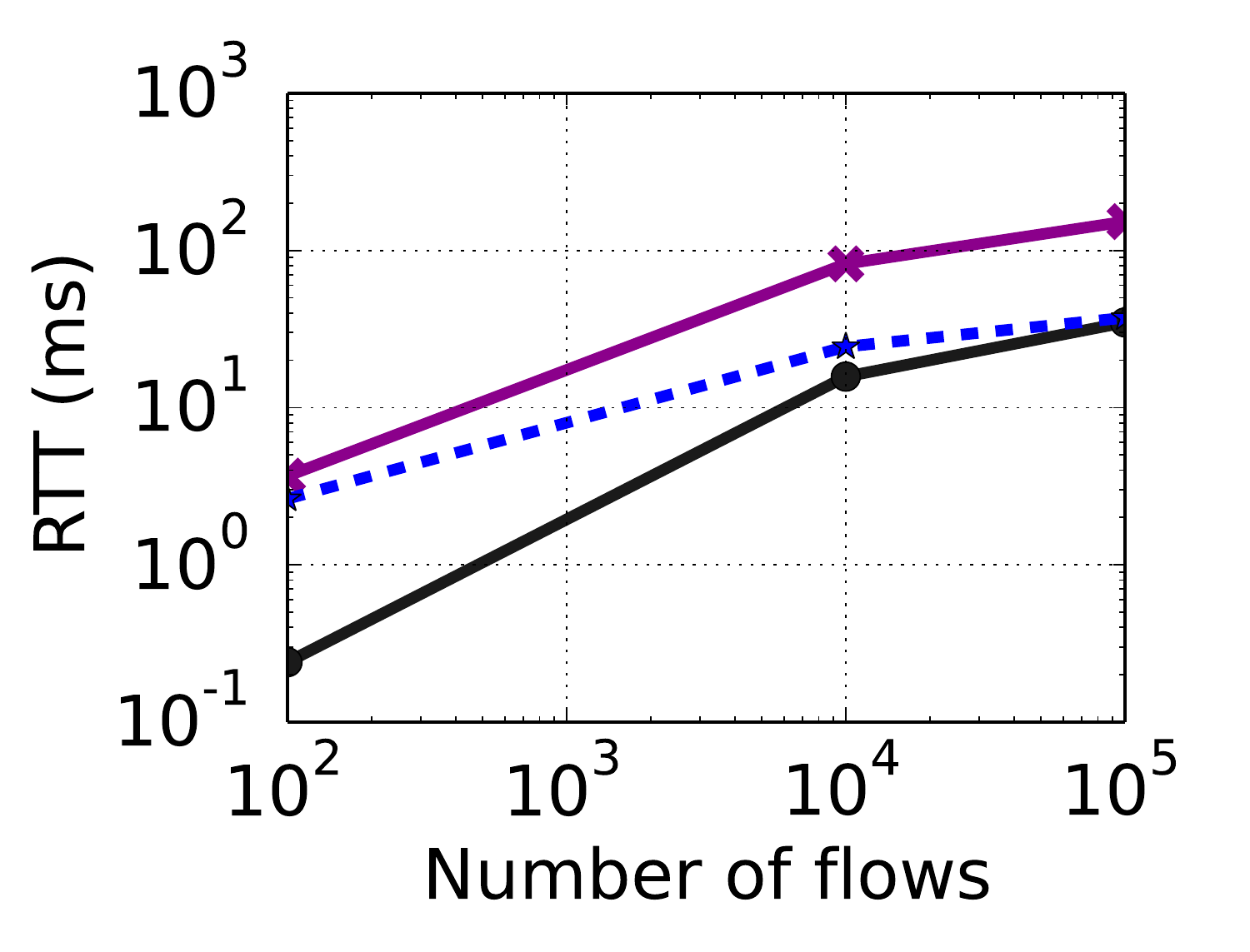}
    \vspace{-0.25in}
    \caption{RTT}
    \label{fig:summary-rtt-9000}
    \end{subfigure}
    \begin{subfigure}{0.48\textwidth}
    \centering
    \includegraphics[width=\linewidth]{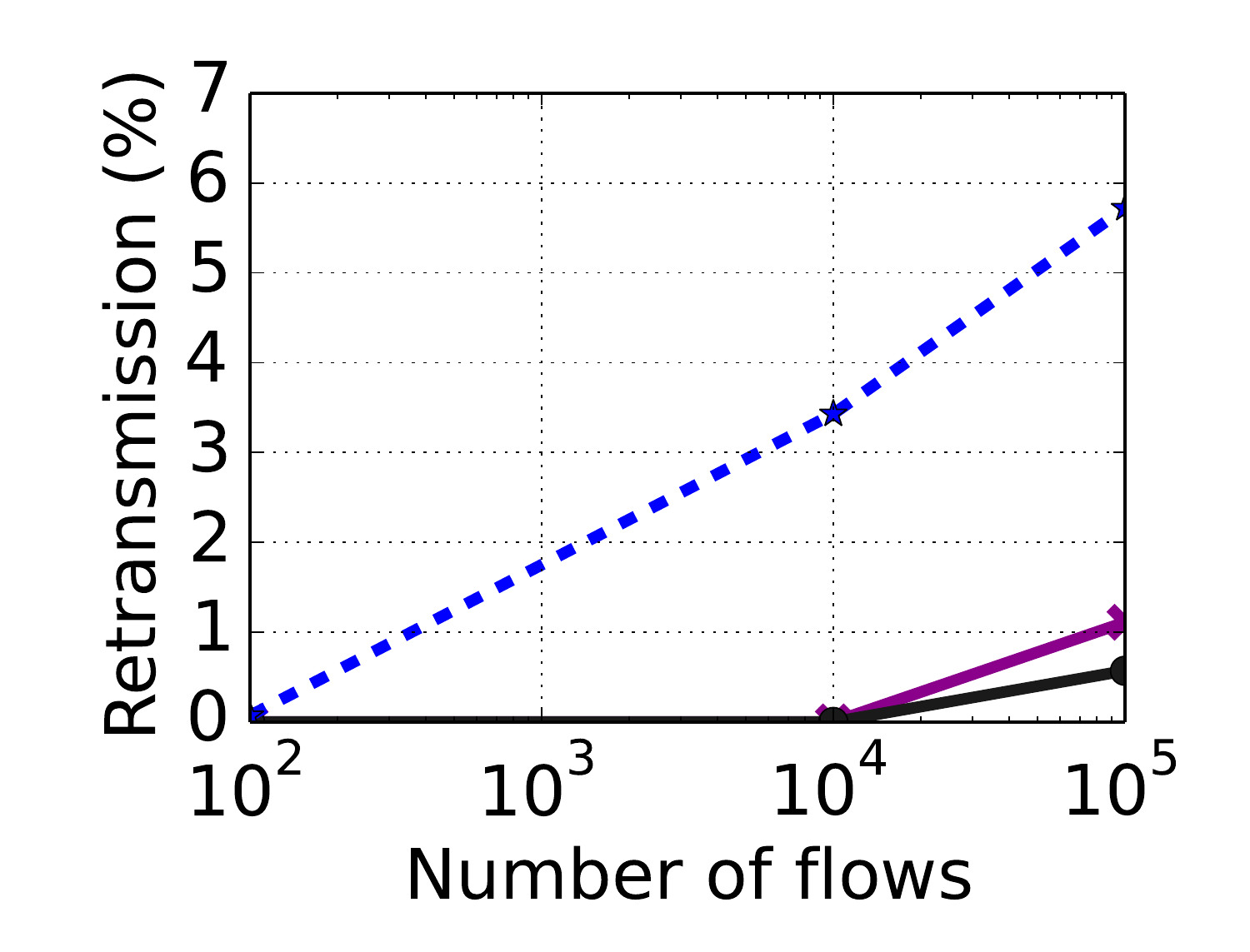}
    \vspace{-0.25in}
    \caption{Retransmission}
    \label{fig:summary-retrans-9000}
    \end{subfigure}
\vspace{-0.1in}
\caption{Overall performance of the network stack as a function of the number of flows with TSO enabled and 9000 MTU size}
\label{fig:summary-9000}
\end{minipage}
\vspace{-0.3in}
\end{figure}

\section{Parameter Configuration}
\label{apx:config}

Table~\ref{tb:tuned-parameter} shows all the parameters we have used in our setup.

\vspace{-0.2in}
\begin{table}[h]
\begin{center}
\begin{tabular}{l|l}
\hline
Parameter                        & Tuned                        \\ \hline
RX-Ring                          & MAX {[}4096{]}               \\
net.core.netdev\_max\_backlog    & 65536                        \\
net.core.tcp\_max\_syn\_backlog  & 65536                        \\
net.ipv4.tcp\_rmem               & 8192 65536 16777216          \\
net.ipv4.tcp\_wmem               & 8192 87380 16777216          \\
net.ipv4.tcp\_mem                & 768849       1025133 1537698 \\
net.core.somaxconn               & 65535                        \\
net.netfilter.nf\_conntrack\_max & 600000                       \\
TSO,GSO                          & enabled                      \\
interrupt moderation             & enabled                      \\
irqbalance                       & disabled                     \\ \hline
\end{tabular}
\end{center}
\caption{\small Tuning parameters}
\label{tb:tuned-parameter}
\vspace{-0.4in}
\end{table}

\section{Overall Stack Performance} \label{apt:overall-perf}

We find that the trends shown in Figure~\ref{fig:summary} remain the same regardless of packet rate. In particular, we disable TSO, forcing the software stack to generate MTU packets. This ensures that the packet rate remains relatively constant across experiments. Note that we perform experiments with a maximum number of 100k flows. We try two values for the MTU: 1500 Bytes and 9000 Bytes. As expected, the performance of the server saturates at a much lower number of flows when generating packets of 1500 Bytes (Figure~\ref{fig:summary-1500}). This is because the packet rate increases compared to the experiments discussed in Section~3. One the other hand, the performance of the server when using 9000 Byte packets is similar to that discussed in Section~3 (Figure~\ref{fig:summary-9000}).

\section{FQ v.s. PFIFO} \label{apx:enqueue-latency}
\begin{figure}[h]
\centering
\vspace{-0.1in}
   \begin{minipage}{0.33\textwidth}
     \centering
     \includegraphics[width=\linewidth]{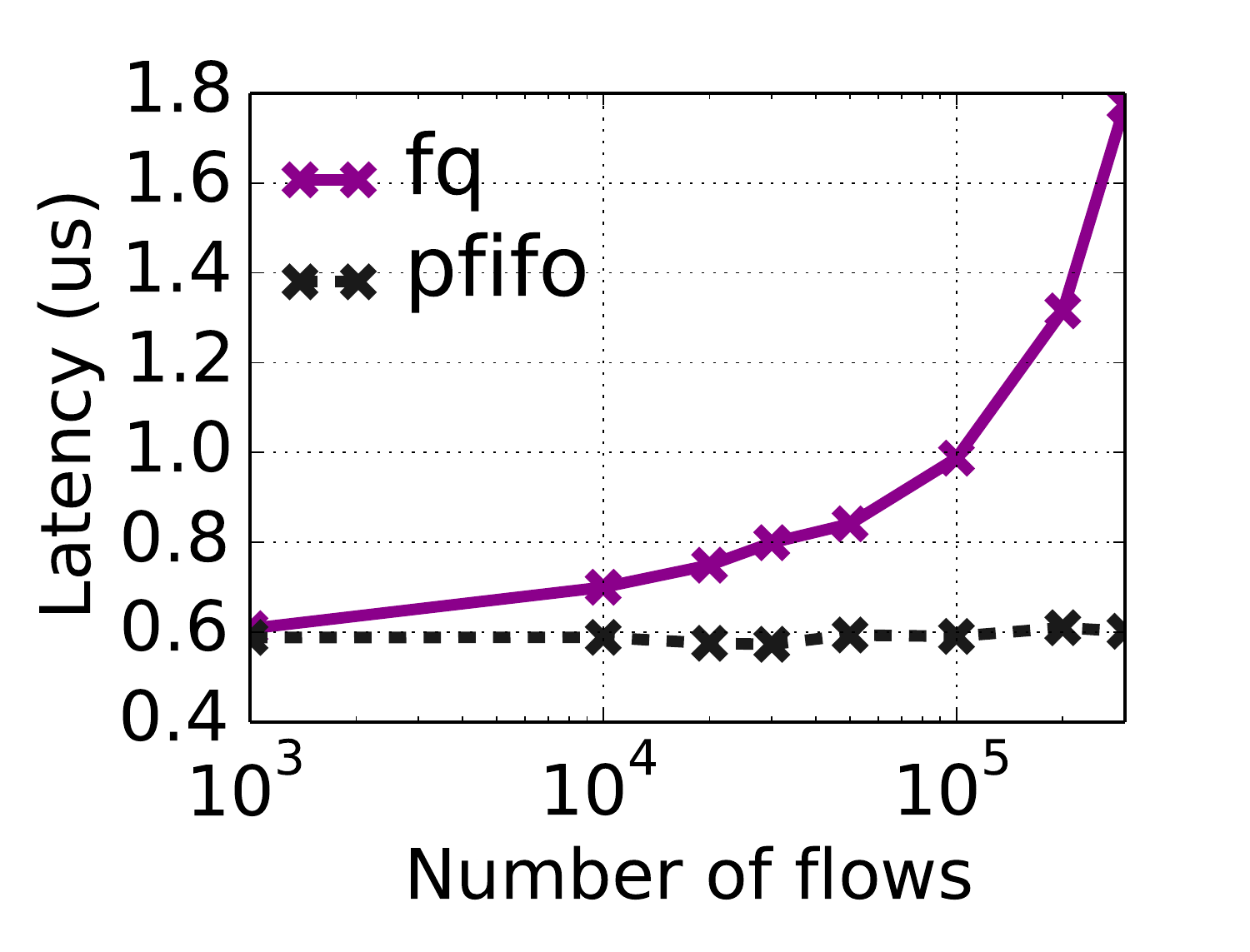}
     \caption{Enqueue time} \label{fig:enq_time}
   \end{minipage}
   \hspace{0.67in}
     \begin{minipage}{0.33\textwidth}
     \centering
     \includegraphics[width=\linewidth]{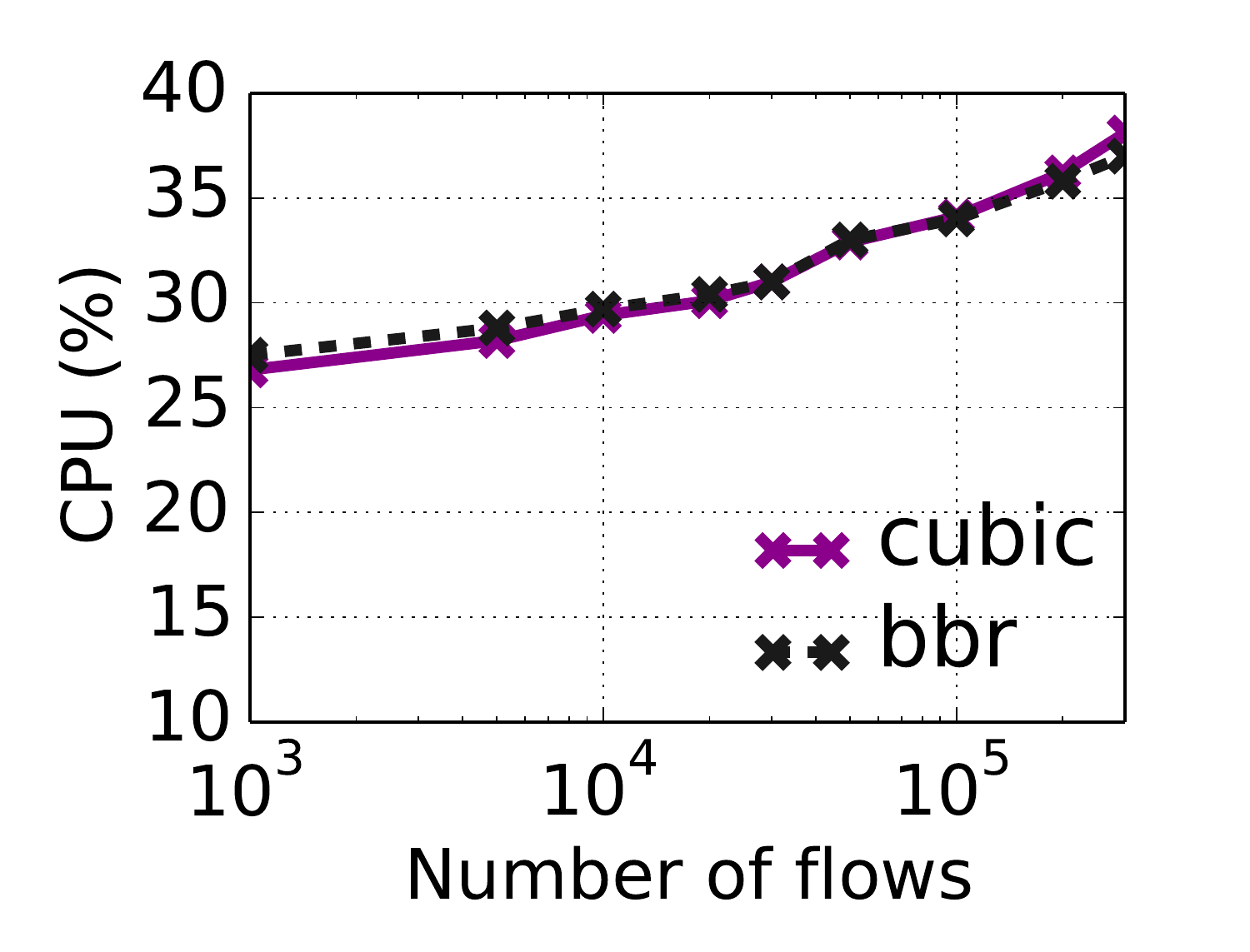}
     \caption{BBR v.s. CUBIC}\label{fig:disable-tso}
   \end{minipage}
\end{figure}

We compare the \verb|fq| with \verb|pfifo_fast| qdiscs in terms of enqueueing latency (Figure \ref{fig:enq_time}). The time to enqueue a packet into \verb|pfifo_fast| queue is almost constant while the enqueue time for \verb|fq| increases with the number of flows. This is because the FQ uses a tree structure to keep track of every flow and the complexity of insertion operation is $O(\log(n))$. The cache miss when fetching flow information from the tree also contributes to the latency with large number of flows.

\section{Packet Rate with Zero Drops}
\label{app:zero_drop}
We verified that BBR and CUBIC has similar CPU usage when PPS is fixed (Figure \ref{fig:disable-tso}). We disable TSO and GSO to fix the packet size and set MTU size to 7000 to eliminate CPU bottleneck. We also observe that with more than 200k flows, CUBIC consumes slightly more CUBIC than BBR because CUBIC reacts to packet drop by reducing packet size, thus generating more packets. 


\end{document}